\documentclass[12pt]{amsart}
     
\usepackage{amssymb}
\usepackage{subfig}
\usepackage{epsfig,amssymb,amsfonts,amsmath,graphicx,dsfont,cite,xfrac}
\usepackage{bbm}
 \usepackage[normalem]{ulem}
\usepackage{float}

\usepackage{cite}

\usepackage{color}
\usepackage{xcolor}
\usepackage[breaklinks,colorlinks,backref]{hyperref}
\hypersetup{
    colorlinks, 
    linktoc=all, 
    linkcolor=red,  
  citecolor=hyptxt,
  urlcolor=blue}
  \hypersetup{
  citebordercolor=,
  filebordercolor=green,
  linkbordercolor=blue
}
\definecolor{hyptxt}{rgb}{0.7, 0.4, 0.9}
\definecolor{mygray}{gray}{0.5}
\usepackage{bibtopic}

\newcommand{\sI}{\mathbbm{1}}

\newcommand{\be}{\begin{equation}}
\newcommand{\ee}{\end{equation}}
\newcommand{\bea}{\begin{eqnarray}}
\newcommand{\eea}{\end{eqnarray}}
\newcommand{\ii}{\mathsf{i}}
\def\ud{\mathrm{d}}

\def\N{\mathbb{N}}

\begin{document}
\date{\today}
 
\title[Generalized Susskind-Glogower coherent states]{Generalized Susskind-Glogower coherent states}
\author[Gazeau-Hussin-Moran-Zelaya]{
Jean-Pierre Gazeau$^{\mathrm{a}}$, V\'eronique Hussin$^{\mathrm{b},\mathrm{c}}$, James Moran$^{\mathrm{b},\mathrm{d}}$, and Kevin Zelaya$^{\mathrm{b}}$}

\address{\emph{$^{\mathrm{A}}$ Universit\'e de Paris, CNRS}\\ \emph{Astroparticule et Cosmologie} \emph{F-75006 Paris, France}}


\address{\emph{$^{\mathrm{B}}$ Centre de Recherches Math\'ematiques}\\ \emph{Universit\'e de Montr\'eal, Montr\'eal QC H3C 3J7, Canada}}

\address{\emph{$^{\mathrm{C}}$ D\'epartement de Math\'ematiques et de Statistique}\\ \emph{Universit\'e de Montr\'eal, Montr\'eal QC H3C 3J7, Canada}}

\address{\emph{$^{\mathrm{D}}$ D\'epartement de Physique}\\ \emph{Universit\'e de Montr\'eal, Montr\'eal QC H3C 3J7, Canada}}

\email{gazeau@apc.in2p3.fr, veronique.hussin@umontreal.ca, james.moran@umontreal.ca, zelayame@crm.umontreal.ca}

{\abstract{
Susskind-Glogower coherent states, whose Fock expansion coefficients include Bessel functions, have recently attracted considerable attention for their optical properties. Nevertheless, identity resolution is still an open question, which is an essential mathematical property that defines an overcomplete basis in the Fock space and allows a coherent state quantization map. In this regard, the \textit{modified Susskind-Glogower} coherent states have been introduced as an alternative family of states that resolve the identity resolution. In the present manuscript, the quantization map related to the modified Susskind-Glogower coherent states is exploited, which naturally leads to a particular representation of the $\mathfrak{su}(1,1)$ Lie algebra in its discrete series. The latter provides evidence about further generalizations of coherent states, built from the Susskind-Glogower ones by extending the indexes of the Bessel functions of the first kind and, alternatively, by employing the modified Bessel functions of the second kind. In this form, the new families of \textit{Susskind-Glogower-I} and \textit{Susskind-Glogower-II} coherent states are introduced. The corresponding quantization maps are constructed so that they lead to general representations of elements of the $\mathfrak{su}(1,1)$ and $\mathfrak{su}(2)$ Lie algebras as generators of the SU$(1,1)$ and SU$(2)$ unitary irreducible representations respectively. For completeness, the optical properties related to the new families of coherent states are explored and compared with respect to some well-known optical states.}}

\maketitle

\tableofcontents

\section{Introduction}
The origins of coherent states trace back to original work by Schr\"odinger~\cite{Sch26}, introduced as wavepackets for the quantum harmonic oscillator so that their maximum point follows a classical trajectory. Nonetheless, the term coherent states (CS) was coined until the works of Glauber and Sudarshan in 1963 within the Fock space of states of the quantum electromagnetic field~\cite{Gla63,Sud63}. Since then, a world of studies have been devoted to these ``standard'' CS on various levels, e.g. algebraic,  analytic, statistical properties~\cite{Man65,Man79}, role played in quantization methods~\cite{Kla00,Gaz09}, in quantum measurement and positive operator-valued measures (POVM)~\cite{Per04,Gaz15}, quantum entanglement~\cite{Kim02}. Furthermore, several generalizations in many directions have been proposed over all these years, starting from spin and atomic CS~\cite{Rad71,Are72}, followed by CS for groups of algebras of many types~\cite{Per86,Wod85,Ros16}, nonlinear CS~\cite{Ber93,Mat96,Man97,Jun99,Zel17,Moj18}, generalization to squeezed states of light~\cite{Wal83,Lou87,Tei89,Zel18}, and recently CS for non-Hermitian~\cite{Zel18A,Gue18,Dey18,Bag20} structures.

Among the several families of nonlinear coherent states, the Susskind-Glogower coherent states~\cite{Rec08} have shown to be a source of optical states with interesting features from a physical point of view \cite{Mon11a,Mon11b}. Such states involve Bessel functions in the coefficients of their linear expansion in the Fock Hilbert space. Despite its success, the proof for identity resolution is, to the best of authors' knowledge, still an open problem. The latter is an essential feature to define an overcomplete basis in the Fock Hilbert space, which is also required for further implementations such as the CS integral quantization map~\cite{Gaz19}. To overcome this issue, a modification to the coefficients of the linear combination of the Susskind-Glogower CS has been introduced in~\cite{CurXX} so that the required identity resolution arises from the summation rules satisfied by the Bessel functions of the first kind. These new states are known as \textit{modified Susskind-Glogower coherent states} and have revealed some interesting nonclassical features, measured through the Mandel parameter and the Helstrom Bound~\cite{CurXX}.

In this manuscript, we start exploiting the CS integral quantization related to the modified Susskind-Glogower CS. This leads to a set of linear operators that become the $\mathfrak{su}(1,1)$ generators associated with the lowest value of the discrete series representation of SU$(1,1)$. The latter suggests the existence of more general families of coherent states that satisfy similar algebraic properties. Thus, we propose two alternative modifications to the Susskind-Glogower CS so that the resulting linear combinations are both normalized and allow the identity resolution. In this form, we arrive at new families of coherent states that generalize the SU$(1,1)$ CS and SU$(2)$ CS. Moreover, the respective quantization maps yield representations of elements of the $\mathfrak{su}(1,1)$ (resp. $\mathfrak{su}(2)$) Lie algebra as generators of the SU$(1,1)$ (resp. SU$(2)$) unitary irreducible representations (UIR) in its discrete series (resp. $2j+1$-dimensional UIR). From the physical perspective, these new families of coherent states reveal useful statistical properties, as well as a highly nonclassical behavior in the context of quantum optics.

The paper is organized as follows. In Section~\ref{sec:SGCS}, we briefly summarize the construction of the Susskind-Glowover and the modified Susskind-Glogower coherent states. The overcompleteness of the modified Susskind-Glogower coherent states is then exploited in Section~\ref{sec:QmSG} to define a quantization map relating optical phase-space functions to unique quantum operators that generate a particular  representation of the $\mathfrak{su}(1,1)$ algebra. In Sections~\ref{sec:C1SG}-\ref{subsec:QmCI}, we introduce a generalization of the modified Susskind-Glogower coherent states achieved by adjusting the index of the Bessel function in the expansion coefficients. Each choice of index defines a new family of \textit{Susskind-Glogower-I coherent states} (SGI CS), from which we explore the quantization mechanism. In this form, we arrive at ladder operators that generate different representations of the $\mathfrak{su}(1,1)$ algebra. For comparison, in Section~\ref{subsec:onephotonCS} we recapitulate some details of the SU$(1,1)$ Perelemov coherent states generated by the action of their unitary displacement operator. In Section~\ref{sec:C2SG} we explore an alternative generalization such that, after replacing the Bessel functions of the first kind with modified Bessel functions of the second kind, we build a family of overcomplete states in finite Hilbert spaces that we call \textit{Susskind-Glogower-II coherent states} (SGII CS). In Section~\ref{subsec:QMCII}, the quantization map using the SGII CS reveals that the generators of the $\mathfrak{su}(2)$ algebra in its spin representations are obtained. 

In Section~\ref{sec:Algebras}, it is shown that, under the appropriate limits, the SGI and SGII CS both contract to reproduce the Glauber-Sudarshan coherent states. Section~\ref{sec:stats} is devoted to comparing the statistical and non-classical properties of the SGI CS with the SU$(1,1)$ CS and the Susskind-Glogower coherent states, and the SGII CS with the SU$(2)$ coherent states. We particularly study the behavior of the Mandel parameter and the physical quadrature variances to look for any trace of squeezing. Finally, In Section~\ref{sec:Conclu} we give our conclusions and some perspectives. For completeness, in Appendix~\ref{sec:norm}, we provide details about the computation of the normalization function related to the SGI CS.

\section{Modified Susskind-Glogower coherent states}
\label{sec:SGCS}
In the context of nonlinear $f$-deformed ladder operators~\cite{Man97}, let us consider the set of operators $\{ \mathbf{V}, \mathbf{V}^{\dagger}\}$ introduced by Susskind-Glogower~\cite{Sus64}, and recently studied in~\cite{Mon11a,Mon11b}, which are defined in terms of the conventional boson and number operators $\mathbf{a}$, $\mathbf{a}^{\dagger}$, $\mathbf{n}=\mathbf{a}^{\dagger}\mathbf{a}$, respectively, through the relation
\begin{equation}
\mathbf{V}:=\sum_{n=1}^{\infty}\vert n-1 \rangle\langle n \vert\equiv \frac{1}{\sqrt{\mathbf{n}+1}}\mathbf{a} \, , \quad \mathbf{V}^{\dagger}:=\sum_{n=0}^{\infty}\vert n+1 \rangle\langle n \vert \equiv \mathbf{a}^{\dagger} \frac{1}{\sqrt{\mathbf{n}+1}} \, ,
\label{operatorV}
\end{equation}
where $(\mathbf{n+1})^{-1/2}\equiv (\mathbf{n}+1)^{-1/2}$. From now on, the identity operator $\mathbbm{1}$ is omitted each time it multiplies a constant. Moreover, $\mathbb{1}$ is the identity operator in the Fock Hilbert space $\mathcal{H}=\overline{\mathrm{Span}\{\vert n \rangle \}_{n=0}^{\infty}}$, i.e. the closure of all finite linear combinations  of the number states $\vert n \rangle$, also known as Fock states~\cite{Dir35}. The operators $\mathbf{V}$ and $\mathbf{V}^{\dagger}$ satisfy the commutation relation $[\mathbf{V},\mathbf{V}^{\dagger}]=\vert 0 \rangle\langle 0 \vert$, which is a projector onto the vacuum state. Such a property has been exploited to construct an exponential operator $D_{\mbox{\tiny SG}}(x)=e^{x(\mathbf{V}^{\dagger}-\mathbf{V})}$, with $x\in\mathbb{R}$, whose action on the vacuum state $\vert 0 \rangle$ leads to a nonlinear family of unit-norm coherent states known as \textit{Susskind-Glogower coherent states}~\cite{Rec08}, given by
\begin{equation}
\vert\alpha\rangle_{\mbox{\tiny SG}}=D_{\mbox{\tiny SG}}(\alpha)\vert 0 \rangle = \sum_{n=0}^{\infty}\alpha^{n}(n+1)\frac{J_{n+1}(2r)}{r^{n+1}}\vert n \rangle \, , \quad \alpha\in\mathbb{C} \, , \quad r:=\vert\alpha\vert \, .
\label{SGcs}
\end{equation}
Above $J_{n}(z)$ the \textit{Bessel functions of the first kind}~\cite{Nik88}, defined through the power series
\begin{equation}
J_{\nu}(z):=\left(\frac{z}{2}\right)^{\nu}\sum_{m=0}^{\infty}\frac{(-1)^{m}\left(\frac{z}{2}\right)^{2m}}{m!\Gamma(\nu+m+1)}  \, .
\label{besselJ}
\end{equation}
The Susskind-Glogower CS have been extensively discussed in the literature, where the respective nonclassical properties have been studied and documented. See~\cite{Mon11a,Mon11b} for details. Nevertheless, to the best of the authors knowledge, the identity resolution is still an open problem, that is, the existence of a weight function $w(\alpha)$ such that
\begin{equation}
\sI:=\sum_{n=0}^{\infty}\vert n\rangle\langle n\vert=\int_{\alpha\in\mathbb{C}}\frac{\ud^{2}\alpha}{\pi}w(r)\vert\alpha\rangle_{\mbox{\tiny SG}}{}_{\mbox{\tiny SG}}\langle\alpha\vert \, ,
\end{equation}
is unknown or might not exist. This implies that the identity resolution generated by the set $\{ \vert \alpha\rangle_{\mbox{\tiny SG}}\}_{\alpha\in\mathbb{C}}$ cannot be taken for granted. It is not known either if this set defines a continuous frame in the sense given in \cite{AliXX}. A workaround for this issue was addressed in~\cite{Gaz19,CurXX}, where the authors have introduced the \textit{modified Susskind-Glogower} coherent states, obtained after modifying the expansion coefficients of the coherent states~\eqref{SGcs} as
\begin{equation}
\vert\alpha\rangle_{\mbox{\tiny mSG}}= \sum_{n=0}^{\infty}\alpha^{n}h_{n}(r)\vert n \rangle \, , \quad h_{n}(r)=\sqrt{\frac{n+1}{\mathcal{N}(r)}}\frac{J_{n+1}(2r)}{r^{n+1}} \, , \quad \alpha\in\mathbb{C} \, .
\label{mSGcs}
\end{equation}
The states $\vert\alpha\rangle_{\mbox{\tiny mSG}}$ have unit norm, while the identity resolution 
\begin{equation}
\sI=\int\frac{\ud^{2}\alpha}{\pi}w(r)\vert\alpha\rangle_{\mbox{\tiny mSG}}{}_{\mbox{\tiny mSG}}\langle\alpha\vert \, ,
\label{IdmSG}
\end{equation}
is fulfilled with the weight function $w(r)=\mathcal{N}(r)$, where $\mathcal{N}(r)$ stands for the normalization constant (for a general proof, see App.~\ref{sec:norm})
\begin{equation}
\mathcal{N}(r):=\frac{1}{r}\sum_{n=0}^{\infty}n [J_{n}(2r)]^{2}={}_{1}F_{2}\left(\left.\begin{aligned} 1/2 \, \\ 2,2 \,\end{aligned}\right\vert -4r^{2}\right) \, .
\label{NmSGcs}
\end{equation}
Such a result ensures that $\{\vert\alpha\rangle_{\mbox{\tiny mSG}} \}_{\alpha\in\mathbb{C}}$ forms an overcomplete  basis in the Fock space $\mathcal{H}$.

Finally, the relation between the elements of the two CS families is given as
\begin{equation}
\label{SGmSG}
\vert\alpha\rangle_{\mbox{\tiny mSG}}  =\frac{1}{\sqrt{\mathcal{N}(\vert \alpha\vert)}} \frac{1}{\sqrt{\mathbf{n}+1}}\vert\alpha\rangle_{\mbox{\tiny SG}} \,.
\end{equation}
Hence, apart from the normalization factor, the modified Susskind-Glogower family results from the action of the compact operator $1/\sqrt{\mathbf{n}+ 1}$ on the Susskind-Glogower family.

\section{Quantization map related to the modified Susskind-Glogower coherent states}
\label{sec:QmSG}
The identity resolution~\eqref{iden-opCS} becomes an essential feature required to construct any quantization mechanism, for it allows to relate an \textit{optical function} $f(\alpha):\mathbb{C}\rightarrow\mathbb{C}$ to a unique linear operator $\mathbf{A}_{f}:\mathcal{H}\rightarrow\mathcal{H}$. Such a procedure is known as the \textit{quantization map}~\cite{Gaz19}, and it is defined in terms of the modified Susskind-Glogower coherent states through the integral transform
\begin{equation}
f(\alpha) \, \mapsto \mathbf{A}_{f}=\int_{\mathbb{C}}\frac{\ud^{2}\alpha}{\pi}w_{\kappa}(r) f(\alpha) \vert\alpha\rangle_{\mbox{\tiny mSG}} \, {}_{\mbox{\tiny mSG}}\langle\alpha\vert \, , \quad r=\vert\alpha\vert \, ,
\label{quant-map}
\end{equation}
so that the quantization of the function $f(\alpha)=1$ leads to the identity operator, a basic requirement in any quantization mechanism~\cite{Gaz19} (see also~\cite{Gos16}). Interestingly, the optical functions $f_{1}(\alpha)=\alpha$ and $f_{2}(\alpha)=\overline{\alpha}$, with $\overline{\alpha}$ the complex-conjugate of $\alpha$, lead to the $f$-deformed ladder operators
\begin{equation}
\alpha\mapsto \mathbf{a}^{(1)}:=\sum_{n=0}^{\infty}a^{(1)}_{n}\vert n \rangle\langle n+1 \vert \, , \quad \overline{\alpha}\mapsto\left(\mathbf{a}^{(1)}\right)^{\dagger}=\sum_{n=0}^{\infty}\overline{\left(a_{n}^{(1)}\right)}\vert n+1 \rangle\langle n\vert \, ,
\label{a1}
\end{equation}
where the matrix elements are
\begin{equation}
a_{n}^{(1)}:=2\sqrt{(n+1)(n+2)}\int_{0}^{\infty}\ud r \, J_{n+1}(2r)J_{n+2}(2r)=\frac{\sqrt{(n+1)(n+2)}}{2} \, .
\label{matrix-a1}
\end{equation}
In the latter result, we have used the identity~\cite{Mag54}
\begin{equation}
\int_{0}^{\infty}\, \ud t \, J_{\nu}(at)J_{\nu+1}(bt)=\begin{cases} a^{\nu}b^{-\nu-1} \quad & 0<a<b \\ \frac{1}{2a} \quad & 0<a=b \\ 0 \quad & 0<b<a \end{cases} \, .
\label{idenBessel1}
\end{equation}
Notice that the following redefined operators:
\begin{equation}
\mathbf{a}^{(1)}_{-}:=2\mathbf{a}^{(1)} \, , \quad \mathbf{a}^{(1)}_{+}:=2\left(\mathbf{a}^{(1)}\right)^{\dagger} \, , \quad 2\mathbf{n}^{(1)}=[\mathbf{a}^{(1)}_{-},\mathbf{a}^{(1)}_{+}]=\mathbf{n}+1 \, ,
\end{equation}
fulfill the commutation relations
\begin{equation}
[\mathbf{n}^{(1)},\mathbf{a}_{+}^{(1)}]=\pm \mathbf{a}_{\pm}^{(1)} \, , \quad [\mathbf{a}_{-}^{(1)},\mathbf{a}_{+}^{(1)}]=2\mathbf{n}^{(1)} \, .
\end{equation}
That is, the quantization map of the optical functions $\alpha$ and $\overline{\alpha}$ leads naturally to a set of linear mappings that serve as generators of the $\mathfrak{su}(1,1)$ Lie algebra in their Fock representation. Moreover, it is well-known that such an algebra admits the \textit{Casimir} operator~\cite{Per86} 
\begin{equation}
\mathbf{C}:=\left(\mathbf{n}^{(1)}\right)^{2}-\frac{1}{2}\left( \mathbf{a}_{+}^{(1)}\mathbf{a}_{-}^{(1)}+\mathbf{a}_{-}^{(1)}\mathbf{a}_{+}^{(1)}\right) \, ,
\label{C1}
\end{equation} 
that commutes with all the generators of the algebra, i.e., $[\mathbf{C},\mathbf{a}_{\pm}^{(1)}]=[\mathbf{C},\mathbf{n}^{(1)}]=0$. Also, the \textit{Bargmann index}~\cite{Ros16}, denoted by $\kappa$, provides information about the unitary irreducible representations of SU$(1,1)$. 

In general, for an orthogonal basis $\{\vert \kappa; m \rangle\}_{m=0}^{\infty}$ carrying the irreducible representation associated with $\kappa$, it follows that $\mathbf{C^{( \kappa)}}\vert \kappa ; m\rangle = \kappa(\kappa-1)\vert\kappa;m\rangle$ holds for any $m=0,1,\cdots$. In the present situation, with the Casimir operator~\eqref{C1} and the Fock Hilbert space $\mathcal{H}$, we have
\begin{equation}
\mathbf{C}\vert n \rangle = 0 \vert n \rangle =0\, , \quad n=0,1,\cdots \, .
\end{equation}
Hence, $\mathbf{C}\equiv \mathbf{C^{(1)}}$ and the quantization map~\eqref{a1} is related to a realization with Bargmann index $\kappa=1$. 

In a previous work~\cite{Olm20}, in the context of the quantization map, it was shown that the SU$(1,1)$ CS associated with the lowest value of the Bargmann index for the square integrable holomorphic discrete series, $\kappa=1$, lead to problems whenever some basic optical functions are being quantized. Nevertheless, the results obtained in this section reveal an intrinsic relation between the modified Susskind-Glogower coherent states and the $\mathfrak{su}(1,1)$ algebra that deserves to be explored in full detail. Such a task is done in the next sections.

\section{Susskind-Glogower-I coherent states}
\label{sec:C1SG}
The latter section provides evidence that a further generalization of the modified Susskind-Glogower should be feasible so that realization of the $\mathfrak{su}(1,1)$ algebra can be achieved for arbitrary values of the Bargmann index. In this section, we explore such a generalization by introducing an arbitrary parameter $\kappa$ into the index of the Bessel functions of the modified Susskind-Glogower coherent states, that is, we consider the change $J_{n+1}(2r)\rightarrow J_{n+\kappa}(2r)$ in ~\eqref{mSGcs}. Thus, we introduce the family of \textit{Susskind-Glogower-I coherent states} (SGI CS) constructed through the linear combination
\begin{equation}
\vert\alpha;\kappa\rangle_{\operatorname{I}}:= \sum_{n=0}^{\infty}\alpha^{n}h_{n;\kappa}(r)\vert n \rangle \, , \quad h_{n;\kappa}(r)=\sqrt{\frac{\mathcal{C}_{n;\kappa}}{\mathcal{N}_{\kappa}(r)}}\frac{J_{n+\kappa}(2r)}{r^{n+\kappa}} \, , \quad r=\vert\alpha\vert \, ,
\label{opSGcs}
\end{equation}
where $\mathcal{N}_{\kappa}(r)$ stands for the normalization constant, and $\mathcal{C}_{n;\kappa}$ are unknown coefficients to be determined. The linear combination~\eqref{opSGcs} must satisfy two basic properties, normalizability for $\alpha\in\mathbb{C}$ and identity resolution. First, we focus on the identity resolution so that the unknown coefficients $\mathcal{C}_{n;k}$ are uniquely defined. Such a task is achieved by following the identity~\cite{Gra07}
\begin{equation}
\int_{0}^{\infty} \ud t\, J_{\nu}(\alpha t)J_{\mu}(\alpha t) t^{-\lambda}=\frac{\alpha^{\lambda-1}\Gamma(\lambda)\Gamma\left(\frac{\nu+\mu-\lambda+1}{2}\right)}{2^{\lambda}\Gamma\left(\frac{-\nu+\mu+\lambda+1}{2}\right)\Gamma\left(\frac{\nu-\mu+\lambda+1}{2}\right)\Gamma\left(\frac{\nu+\mu+\lambda+1}{2}\right)} \, ,
\label{idenBessel0}
\end{equation}
which converges for Re$(\nu+\mu+1)>\mathrm{Re}\lambda>0$ and $\alpha>0$. Thus, the identity resolution associated with the linear combination~\eqref{opSGcs} is determined from
\begin{equation}
\sI=\int_{\alpha\in\mathbb{C}}\frac{\ud^{2}\alpha}{\pi}w_{\kappa}(r)\vert \alpha;\kappa\rangle_{\operatorname{I}}\, {}_{\operatorname{I}}\langle\alpha;\kappa\vert=\sum_{n=0}^{\infty}2\mathcal{C}_{m;\kappa}\mathcal{D}_{\kappa}\int_{0}^{\infty}\ud r \, r^{-(2\kappa-1)}\left[ J_{n+\kappa}(2r)\right]^{2} \vert n \rangle\langle n \vert \, ,
\label{iden-opCS}
\end{equation}
with the weight function fixed as $w_{\kappa}(r)=\mathcal{D}_{\kappa}\mathcal{N}_{k}(r)$, and  $\mathcal{D}_{\kappa}$ a proportionality factor independent of $n$. The integral in~\eqref{iden-opCS} is solved by means of the identity~\eqref{idenBessel0}, leading to
\begin{equation}
\sI=\sum_{n=0}^{\infty}\mathcal{C}_{n;\kappa}\mathcal{D}_{\kappa}\frac{2^{2\kappa-1}(1/2)_{\kappa}}{(2\kappa-1)\Gamma(\kappa)}\frac{1}{(n+1)_{2\kappa-1}} \vert n \rangle\langle n \vert \, ,
\end{equation}
where $(a)_{\nu}:=\Gamma(a+\nu)/\Gamma(a)$ stands for the Pochhammer symbol~\cite{Olv10}. From the latter,  the identity resolution is achieved after fixing
\begin{equation}
\mathcal{C}_{n;\kappa}=(n+1)_{2\kappa-1}=\frac{\Gamma(2\kappa+n)}{n!} \, , \quad \mathcal{D}_{\kappa}=\frac{(2\kappa-1)\Gamma(\kappa)}{2^{2\kappa-1}(1/2)_{\kappa}} \, .
\label{Ck}
\end{equation}
From the convergence conditions of the integral~\eqref{idenBessel0}, we conclude that the matrix elements~\eqref{iden-opCS} converge for all $n=0,1,\cdots$ provided that $\kappa>\frac{1}{2}$. From~\eqref{Ck}, we have completely characterized the SGI CS given in~\eqref{opSGcs}. Notice that the coefficients $\mathcal{C}_{n;\kappa}$ correspond to the weighting factors of a negative binomial distribution. The latter is also a common property of the SU$(1,1)$ Perelomov CS~\cite{Per86,Gaz19}, but with different weighting factors. Thus, it seems that indeed both families of coherent states are related in a way. More details are discussed in the sequel.

Now, we must verify the normalizability of the SGI CS.  Interestingly, an analytic expression can be determined for $\mathcal{N}_{\kappa}$, which is computed after expanding the Bessel functions in power series, and after arranging the resulting sums in a convenient form. See App.\ref{sec:norm} for a detailed proof. We thus have
\begin{equation}
\mathcal{N}_{\kappa}(r):=\frac{\Gamma(2\kappa)}{[\Gamma(\kappa+1)]^{2}} \,\,{}_{1}F_{2}\left(\left.\begin{aligned} &1/2 \\ \kappa+1&,\kappa+1 \end{aligned}\right\vert -4r^{2}\right) \, ,
\label{Normk}
\end{equation}
with 
\begin{equation}
{}_{p}F_{q}\left(\left.\begin{aligned} a_{1},\cdots a_{p} \\ b_{1},\cdots b_{q}\end{aligned}\right\vert z \right)=\sum_{n=0}^{\infty}\frac{(a_{1})_{n}\cdots (a_{p})_{n}}{(b_{1})_{n}\cdots (b_{q})_{n}} \frac{z^{n}}{n!} \, ,
\label{pFq}
\end{equation}
the \textit{generalized hypergeometric function}~\cite{Olv10}. The series expansion~\eqref{pFq} converges in $z\in\mathbb{C}$ for $q\geq p$, where the generalized hypergeometric function becomes an entire function~\cite{Olv10}. For $p=q+1$, the series converges inside the open unit-disk $\vert z\vert<1$. 

Therefore, the SGI CS constructed by the linear combination $\vert \alpha;\kappa\rangle=\sum_{n=0}^{\infty}c_{n;\kappa}^{(I)}(r)\vert n \rangle$, with expansion coefficients 
\begin{equation}
c_{n;\kappa}^{(I)}(r)=\sqrt{\frac{(2\kappa)_{n}}{n!}}\,\Gamma(\kappa+1) \left[{}_{1}F_{2}\left(\left.\begin{aligned} 1/2 \hspace{6mm}\, \\ \kappa+1,\kappa+1 \end{aligned}\right\vert\, -4r^{2}\right)\right]^{-1/2}\frac{e^{\ii n\phi}J_{n+\kappa}(2r)}{r^{k}} \, ,
\label{CIcn}
\end{equation}
define an overcomplete and normalizable set $\{\vert\alpha;\kappa\rangle\}_{\alpha\in\mathbb{C}}$. 

\section{Quantization map with SGI CS}
\label{subsec:QmCI}
With the use of the SGI CS, and from the identity resolution~\eqref{iden-opCS}, it is possible to repeat the integral quantization mechanism \eqref{quant-map} that takes a given \textit{optical function} $f(\alpha):\mathbb{C}\rightarrow\mathbb{C}$ and transforms it into an operator $\mathbf{A}_{f}$ in $\mathcal{H}$ as
\begin{equation}
f(\alpha) \, \mapsto \mathbf{A}_{f}=\int_{\mathbb{C}}\frac{\ud^{2}\alpha}{\pi}w_{\kappa}(r) f(\alpha)\vert\alpha;\kappa\rangle_{\operatorname{I}}\, {}_{\operatorname{I}}\langle\alpha;\kappa\vert \, , \quad r=\vert\alpha\vert \, .
\label{quant-map1}
\end{equation}
In analogy to the results found in Sec.~\ref{sec:QmSG}, we consider $f_{1}(\alpha)=\alpha$ and $f_{2}(\alpha)=\overline{\alpha}$.  Since the weight $w_{\kappa}(r)$ is a real-valued function, it is clear that the quantization map of $\alpha\mapsto\mathbf{a}^{(k)}$ and $\overline{\alpha}\mapsto\tilde{\mathbf{a}}^{(\kappa)}$ leads to the mutually adjoint operators $\tilde{\mathbf{a}}^{(\kappa)}\equiv \left(\mathbf{a}^{(\kappa)}\right)^{\dagger}$. In this regard, it is only necessary to compute one of the quantization maps, namely $\mathbf{a}^{(\kappa)}$. The straightforward calculation leads to
\begin{align}
&\alpha \mapsto \mathbf{a}^{(\kappa)}:=\sum_{n=0}a^{(\kappa)}_{n}\vert n\rangle\langle n+1\vert \, , \quad \overline{\alpha} \mapsto \left(\mathbf{a}^{(\kappa)}\right)^{\dagger}=\sum_{n=0}^{\infty}\overline{\left(a^{(\kappa)}_{n}\right)}\vert n+1 \rangle\langle n \vert \, ,
\label{ah0}
\end{align}
where the matrix elements are determined through
\begin{equation}
a_{n}^{(\kappa)}:=2\mathcal{D}_{\kappa}\sqrt{(n+1)(n+2\kappa)}(n+2)_{2\kappa-2}\int_{0}^{\infty}\ud r \, r^{-(2\kappa-2)} J_{n+\kappa}(2r)J_{n+\kappa+1}(2r) \, .
\label{matrix-a}
\end{equation}
With the aid of~\eqref{idenBessel0} we get
\begin{equation}
a_{n}^{(\kappa)}=\frac{\sqrt{(n+1)(n+2\kappa)}}{2} \, , \quad \kappa>1 \, , \quad n=0,1,\cdots \, .
\end{equation}
Contrary to the identity resolution~\eqref{iden-opCS}, where $\kappa>1/2$, the quantization of $\alpha$ and $\overline{\alpha}$ is achieved only for $\kappa>1$. Although $\kappa=1$ is excluded from the previous results, it was already considered in Sec.~\ref{sec:QmSG}. Thus, $\kappa\rightarrow 1$ becomes a valid limit case.

Now, it is convenient to introduce the rescaled operators
\begin{align}
&\mathbf{a}^{(\kappa)}_{-}:=2\mathbf{a}^{\kappa}=\sum_{n=0}^{\infty}\sqrt{(n+1)(n+2\kappa)}\vert n \rangle\langle n+1 \vert \equiv \mathbf{a}\sqrt{\mathbf{n}+2\kappa-1} \, ,
\label{ah} \\
&\mathbf{a}^{(\kappa)}_{+}:=2\left(\mathbf{a}^{\kappa}\right)^{\dagger}=\sum_{n=0}^{\infty}\sqrt{(n+1)(n+2\kappa)}\vert n+1 \rangle\langle n \vert \equiv \sqrt{\mathbf{n}+2\kappa-1}\,\mathbf{a}^{\dagger} \, .
\label{ahd}
\end{align}
valid for $\kappa\geq 1$. In this form, we can define a third operator from the commutation relation between the ladder operators, such that we get
\begin{equation}
\mathbf{n}^{(\kappa)}:=\frac{1}{2}[\mathbf{a}^{(\kappa)}_{-},\mathbf{a}^{(\kappa)}_{+}]=\mathbf{n}+\kappa \, , \quad [\mathbf{n}^{(\kappa)},\mathbf{a}^{(\kappa)}_{\pm}]=\pm\mathbf{a}^{(\kappa)}_{\pm} \, ,
\label{realization-su11}
\end{equation}
In this form, the set $\{\mathbf{a}^{(\kappa)}_{-},\mathbf{a}^{(\kappa)}_{+},\mathbf{n}^{(\kappa)}\}$, together with the Casimir operator 
\begin{equation}
\mathbf{C}^{(\kappa)}=\left(\mathbf{n}^{(\kappa)}\right)^{2}-\frac{1}{2}\left(\mathbf{a}^{(\kappa)}_{-}\mathbf{a}^{(\kappa)}_{+}+\mathbf{a}^{(\kappa)}_{+}\mathbf{a}^{(\kappa)}_{-} \right) \, , \quad [\mathbf{C}^{(\kappa)},\mathbf{a}^{(\kappa)}_{\pm}]=[\mathbf{C}^{(\kappa)},\mathbf{n}^{(\kappa)}]=0 \, ,
\end{equation}
reveals that
\begin{equation}
\label{Casimir}
 \mathbf{C}^{(\kappa)}\vert n \rangle=\kappa(\kappa-1)\vert n\rangle\,.
\end{equation} 
That is, the quantization map obtained with SGI CS leads to a set of ladder operators that serve as the representation of $\mathfrak{su}(1,1)$ with Bargmann index $\kappa\geq 1$. Notice that such generators are a \textit{one-mode and one-photon realization} of $\mathfrak{su}(1,1)$, contrary to the conventional two-photon realization~\cite{Nie97} provided by the set $\{\frac{1}{2}\mathbf{a}^{2},\frac{1}{2}\left(\mathbf{a}^{\dagger}\right)^{2},\frac{1}{2}\left(\mathbf{n}+\frac{1}{2}\right) \}$. 

The family of coherent states constructed as eigenstates of $\mathbf{a}^{h}_{-}$, together with its completeness and nonclassical properties, has been reported in~\cite{Zel17}. Nevertheless, those coherent states are not considered throughout this work and, an alternative construction will be considered in the next section.

In quantum optics and within the context of Glauber-Sudarshan coherent states~\cite{Gla63}, the function $\vert\alpha\vert^{2}$ plays the role of the field intensity~\cite{Gaz19} since it is equal to the average number of photons, $\langle\mathbf{n}\rangle = \vert\alpha\vert^{2}$. For generalized nonlinear coherent states,  the average number of photons $\langle\mathbf{n}\rangle$ is no longer proportional to $\vert\alpha\vert^{2}$. This property is discussed in detail in the Sec.~\ref{sec:stats}. Nevertheless, it is interesting to compute here the quantization map of the function $f(\alpha)=\vert\alpha\vert^{\gamma}$, with $\gamma$ a real parameter. We get
\begin{align}
&\vert\alpha\vert^{\gamma} \mapsto \mathbf{A}^{(\kappa;\gamma)}=\int_{\mathbb{C}}\frac{\ud^{2}\alpha}{\pi}w(r)r^{\gamma}\vert\alpha;\kappa\rangle_{\operatorname{I}}\,{}_{\operatorname{I}}\langle\alpha;\kappa\vert=\sum_{n=0}^{\infty}A_{n}^{(\kappa;\gamma)}\vert n \rangle\langle n \vert \, ,	\\
& A_{n}^{(\kappa;\gamma)}:=2\mathcal{D}_{k}(n+1)_{2\kappa-1}\int_{0}^{\infty} \ud r \, r^{-(2\kappa-\gamma-1)} [J_{n+\kappa}(2r)]^{2} \, .
\label{Ann}
\end{align}
Clearly, the latter integral converges for some specific values of $\gamma$ and $\kappa$. For instance, the asymptotic behavior of $J_{\nu}(z)$~\cite{Nik88,Olv10} for $z\ll1$, and $z \gg 1$, leads to
\begin{align}
&\left. r^{-(2\kappa-\gamma-1)}[J_{n+\kappa}(2r)]^{2}\right\vert_{r\rightarrow 0} \sim \frac{x^{2n+\gamma+1}}{[\Gamma(n+\kappa+1)]^{2}} \, , \\
&\left.r^{-(2\kappa-\gamma-1)}[J_{n+\kappa}(2r)]^{2}\right\vert_{r\rightarrow\infty} \sim \frac{2r^{-(2\kappa-\gamma-2)}}{\pi}\cos\left(r-\frac{\pi}{2}\left(n+\kappa+\frac{1}{2}\right)\right) \, ,
\end{align}
respectively. Thus, from the integrand in~\eqref{Ann}, it is clear that the condition $-2<\gamma<2\kappa-1$ is necessary to avoid singularities around $r\rightarrow 0$, and to ensure convergence at $r\rightarrow\infty$. Moreover, from~\eqref{idenBessel0} we obtain
\begin{equation}
A^{(\kappa;\gamma)}_{n}=\mathcal{D}_{\kappa;\gamma}\frac{(n+1)_{2\kappa-1}}{(n+1+\frac{\gamma}{2})_{2\kappa-\gamma-1}} \, , \quad \mathcal{D}_{\kappa;\gamma}=\frac{1}{2^{\gamma}}\left(\frac{2\kappa-1}{2\kappa-\gamma-1}\right)\frac{(\kappa-\frac{\gamma}{2})_{1/2}}{(\kappa)_{1/2}} \, .
\label{Anngamma}
\end{equation}
As a matter of fact, the quantization map of $\vert\alpha\vert^{2}$ is achieved only for $\kappa>3/2$. However, we should not be so much concerned about this fact since  this quantity does not represent  the field intensity. From the general result~\eqref{Anngamma}, it is worth discussing some particular cases. For $\gamma=2\Lambda$, with $\Lambda=0,1,\cdots$, we obtain a degree $2\Lambda$ polynomial of the number operator $\mathbf{n}$ given by
\begin{equation}
\mathbf{A}^{(\kappa;2\Lambda)}=\mathcal{D}_{\kappa;2\Lambda}(\mathbf{n}+1)_{\Lambda}(\mathbf{n}+2\kappa-\Lambda)_{\Lambda} \, , \quad \kappa>\Lambda+\frac{1}{2} \, .
\end{equation}
In such a case, we get nonlinear quantized field amplitudes, from which the linear case $c_{1}\mathbf{n}+c_{0}$, with $c_{0}$ and $c_{1}$ arbitrary constants, cannot be extracted under any special limit. 

On the other hand, for $\gamma=-1$, we have the rational function of the number operator
\begin{equation}
\mathbf{A}^{(\kappa,-1)}=\mathcal{D}_{\kappa,-1}\frac{(\mathbf{n}+1)_{2\kappa-1}}{(\mathbf{n}+1/2)_{2\kappa}} \, , \quad \kappa>1 \, .
\end{equation}
For other values of $\gamma$, the functional form of the operator $\mathbf{A}^{(\kappa,\gamma)}$ cannot be reduced to neither polynomial or rational functions of $\mathbf{n}$.

\section{One-photon SU$(1,1)$ coherent states}
\label{subsec:onephotonCS}
In the previous section, we have found that the quantization map of $\alpha$ and $\overline{\alpha}$ through the coherent states~\eqref{opSGcs} lead naturally to a one-photon realization of the $\mathfrak{su}(1,1)$ Lie algebra. In this section, we deal with the families of coherent states contructed through the action of the displacement operator $D_{\kappa}(z)= \exp\left(z\mathbf{a}^{(\kappa)}_{+}-z^{*}\mathbf{a}^{(\kappa)}_{+}\right)$ on the fiducial state, in this case $\vert 0 \rangle$. The unitary displacement operator can be decomposed by means of the well-known disentanglement formula~\cite{Ger05,Kok10}
\begin{equation}
D_{\kappa}(\tau(z,z^{*}))\equiv e^{\tau\mathbf{a}^{(\kappa)}_{+}}e^{-2\nu\mathbf{n}^{(\kappa)}}e^{-\tau^{*}\mathbf{a}^{(\kappa)}_{-}} \, , \quad \tau=\frac{z}{\vert z\vert}\tanh(\vert z\vert) \ , \quad \nu=\ln\cosh(\vert z\vert) \, ,
\end{equation}
leading to the coherent states
\begin{equation}
\vert \tau;\kappa \rangle_{\operatorname{\mathfrak{su}(1,1)}}:=D_{\kappa}(\tau)\vert 0 \rangle = \left(1-\vert\tau\vert^{2}\right)^{\kappa}\sum_{n=0}^{\infty} \tau^{n}\sqrt{\frac{(2\kappa)_{n}}{n!}}\, \vert n\rangle \, , \quad \vert\tau\vert<1 \, ,
\label{Psu11}
\end{equation}
which satisfy the overlap formula
\begin{equation}
{}_{\operatorname{\mathfrak{su}(1,1)}}\langle\tau_{2};\kappa\vert\tau_{1};\kappa\rangle_{\operatorname{\mathfrak{su}(1,1)}}=\frac{\left(1-\vert\tau_{1}\vert^{2}\right)^{\kappa}\left(1-\vert\tau_{2}\vert^{2}\right)^{\kappa}}{\left(1-\tau_{1}\overline{\tau}_{2}\right)^{2\kappa}} \, ,\quad \vert\tau_{i}\vert<1 \, , \quad i=1,2 \, .
\end{equation}
The latter corresponds to the well-known Perelemov SU$(1,1)$ coherent states~\cite{Per86} defined in the unit-disk and built from   the     UIR  in the discrete series corresponding to  the value \eqref{Casimir} of the Casimir operator determined by the parameter $\kappa\geq 1$ introduced in the SGI CS~\eqref{opSGcs}. In contradistinction to the Barut-Girardello construction, previously reported in~\cite{Zel17}, the states $\vert\tau;\kappa\rangle_{\operatorname{\mathfrak{su}(1,1)}}$ do not belong to the set of eigenstates of the annihilation operator $\mathbf{a}^{(\kappa)}_{-}$, instead we get
\begin{equation}
\mathbf{a}^{(\kappa)}_{-}\vert\tau;\kappa\rangle_{\operatorname{\mathfrak{su}(1,1)}}=2\kappa\tau\vert\tau;\kappa\rangle_{\operatorname{\mathfrak{su}(1,1)}}+\tau^{2}\frac{\partial}{\partial\tau}\left[\left(1-\vert\tau\vert^{2}\right)^{-\kappa}\vert\tau;\kappa\rangle_{\operatorname{\mathfrak{su}(1,1)}}\right] \, .
\end{equation}
The identity resolution for the family of coherent states~\eqref{Psu11} is achieved by means of the weight function $\tilde{w}_{\kappa}(r)$ as 
\begin{equation}
\sI=\int_{\vert\tau\vert<1}\frac{\ud^{2}\tau}{\pi}\tilde{w}_{k}(r) \vert \tau;\kappa \rangle_{\operatorname{\mathfrak{su}(1,1)}}\,{}_{\operatorname{\mathfrak{su}(1,1)}}\langle\tau;\tau\vert \, , \quad \tilde{w}(r)=\frac{2\kappa-1}{(1-r^{2})^{2}} \, .
\label{HPiden}
\end{equation}
Thus, the set $\{\vert\tau;\kappa\rangle_{\operatorname{\mathfrak{su}(1,1)}}\}_{\vert\tau\vert<1}$ forms an overcomplete family for the Hilbert  space $\mathcal{K}_{\kappa}$ of square-integrable functions on the open unit-disk  with measure $\dfrac{\ud^{2}\tau}{\pi}\tilde{w}_{\kappa}(r)$. In analogy to the SSI CS, we can exploit the identity resolution to construct the appropriate quantization map, this time defined on the open unit-disk as
\begin{equation}
g(\tau) \mapsto \mathbf{G}:=\int_{\vert\tau\vert<1}\frac{\ud^{2}\tau}{\pi}\tilde{w}_{\kappa}(r)g(\tau)\,\vert \tau;\kappa\rangle_{\operatorname{\mathfrak{su}(1,1)}}\,{}_{\operatorname{\mathfrak{su}(1,1)}}\langle\tau;\kappa\vert \, .
\end{equation}
In particular, we consider the quantization maps $(1-\vert\tau\vert^{2})^{-1}\tau \mapsto\mathbf{b}$ and $(1-\vert\tau\vert^{2})^{-1}\overline{\tau} \mapsto\mathbf{b}^{\dagger}$, which leads to the annihilation and creation operators
\begin{equation}
\mathbf{b}^{(\kappa)}:=\sum_{n=0}^{\infty}\frac{\sqrt{(n+1)(n+2\kappa)}}{2(\kappa-1)}\vert n\rangle\langle n+1\vert \, , \quad \left(\mathbf{b}^{(\kappa)}\right)^{\dagger}:=\sum_{n=0}^{\infty}\frac{\sqrt{(n+1)(n+2\kappa)}}{2(\kappa-1)}\vert n+1 \rangle\langle n \vert \, ,
\end{equation}
respectively. From the latter, it is clear that the new ladder operators are well-behaved for $\kappa>1$, that is, for the lower value of the discrete series $\kappa=1$ the quantization map is not defined, a fact already noticed in~\cite{Olm20}. Interestingly, after introducing the proper reparametrization $\mathbf{b}_{-}\equiv 2(\kappa-1)\mathbf{b}$ and $\mathbf{b}_{+}\equiv 2(\kappa-1)\mathbf{b}^{\dagger}$, it follows that the set $\{\mathbf{b}_{-},\mathbf{b}_{+},\frac{1}{2}[\mathbf{b}_{-},\mathbf{b}_{+}]\}$ realizes the $\mathfrak{su}(1,1)$ Lie algebra.

Additionally, the  function $(1-\vert\tau\vert^{2})^{-1}\vert\tau\vert^{2\gamma}$ leads to the operator
\begin{equation}
\frac{\vert\tau\vert^{2\gamma}}{1-\vert\tau\vert^{2}}\mapsto \mathbf{B}^{(\kappa;\gamma)}:=\frac{1}{2(\kappa-1)}\sum_{n=0}^{\infty}\frac{(n+1)_{\gamma}}{(n+2\kappa)_{\gamma-1}}\vert n\rangle\langle n\vert \, ,
\label{Bq}
\end{equation}
which, for $\gamma=0$ and $\gamma=1$, reduces to the linear functions of the number operator
\begin{equation}
\mathbf{B}^{(\kappa;0)}\equiv \frac{\mathbf{n}+2\kappa-1}{2(\kappa-1)} \, , \quad \mathbf{B}^{(\kappa;1)}\equiv \frac{\mathbf{n}+1}{2(\kappa-1)} \, .
\end{equation}
Notice that, for $\gamma$ equal to any other integer (either positive or negative), the operator $\mathbf{B}^{(\kappa,\gamma)}$ becomes a rational function of the number operator.


\section{Susskind-Glogower-II coherent states}
\label{sec:C2SG}
In this section, we consider an alternative set of coherent states, which rise as a variation of the ones introduced in Sec.~\ref{sec:C1SG}. This is achieved by modifying the functional coefficients $h_{n,\kappa}(r)$, where $\kappa \in\N^{*}/2$, with $\N^{*}=\{1,2,\cdots\}$, and introducing the modified Bessel functions of the second kind $K_{\nu}(r)$, defined as~\cite{Nik88}
\begin{equation}
K_{\nu}(z):=\frac{\pi}{2}\frac{I_{-\nu}(z)-I_{\nu}(z)}{\sin\pi\nu} \, , \quad I_{\nu}(z):=e^{-\ii \pi\nu/2}J_{\nu}(\ii z) \, ,
\label{BesselK}
\end{equation}
with $I_{\nu}(z)$ the modified Bessel function of the first kind. The function $K_{\nu}(z)$ behave, for the asymptotic value $z\rightarrow\infty$, as $K_{\nu}(z)\sim\sqrt{\pi/(2z)}e^{-z}$. For $z\rightarrow 0$, the modified Bessel function of second kind has a branch point for all $\nu\in\mathbb{C}$~\cite{Olv10}. Moreover, $K_{\nu}(z)$ is analytic in $\mathbb{C}\setminus(-\infty,0]$. In analogy to the functions $h_{n;\kappa}(r)$ of~\eqref{mSGcs}, we introduce the new functions $\mathfrak{h}_{n;\kappa}$ written in terms of $K_{\nu}(z)$. We thus introduce the \textit{Susskind-Glogower-II coherent states} (SGII CS) defined as
\begin{equation}
\vert z;\kappa\rangle_{\operatorname{II}}=\sum_{n=0}^{2\kappa}z^{n}\mathfrak{h}_{n;\kappa}(r)\vert n \rangle \, , \quad \mathfrak{h}_{n;\kappa}(r)=\sqrt{\frac{\mathfrak{C}_{n;\kappa}}{\mathfrak{N}_{\kappa}(r)}}\frac{K_{n-\kappa}(2r)}{r^{n-\kappa}} \, , \quad z\in\mathbb{C} \, , \quad r=\vert z\vert \, ,
\label{mSGII}
\end{equation}
where $\mathfrak{N}_{\kappa}(r)$ stands for the normalization factor, and the coefficients $\mathfrak{C}_{n;\kappa}$ are independent of $z$ and such that the set $\{\vert z;\kappa\rangle\}_{z\in\mathbb{C}}$ fulfills the identity resolution
\begin{equation}
\sI_{2\kappa+1}:=\sum_{n=0}^{2\kappa}\vert n \rangle\langle n\vert=\int_{z\in\mathbb{C}}\frac{\ud^{2}z}{\pi}\mathfrak{w}_{k}(r)\vert z;\kappa\rangle_{\operatorname{II}} \, {}_{\operatorname{II}}\langle z;\kappa\vert \, ,
\label{IdmSGII} 
\end{equation}
with $\mathfrak{w}_{k}(r)$ the respective weight function. In this form, the set $\{\vert z;\kappa\rangle\}_{z\in\mathbb{C}}$ generates the $2\kappa +1$-dimensional Hilbert subspace $\mathcal{H}^{(2\kappa)}=\mathrm{Span}\{\vert n \rangle\}_{n=0}^{2\kappa}\subset\mathcal{H}$. Notice that the coherent states~\eqref{mSGII} are defined through a finite linear combination, for which the normalization constant $\mathfrak{N}_{\kappa}(r)$ converges in the complex-plane as long as $\mathfrak{h}_{n;\kappa}(r)$ is free of singularities for $r\in\mathbb{R}^{+}\cup\{0\}$. From the asymptotic behavior previously discussed, we can guarantee the finite-norm condition for $n=0,1,\cdots,2\kappa$. It is worth mentioning that $\mathfrak{h}_{n;\kappa}(r)$ leads to singularities either at $r=0$ or $r\rightarrow\infty$ for $n=2\kappa+1$. For that reason, we have truncated the linear combination in~\eqref{mSGII} and restricted the values of $\kappa$ to non-negative integers or half-integers.

The identity resolution~\eqref{IdmSGII}, together with $\mathfrak{C}_{n;\kappa}$, are determined from the formula~\cite{Gra07}
\begin{multline}
\int_{0}^{\infty}dt\, t^{-\lambda}K_{\mu}(at)K_{\nu}(at)=\frac{2^{-2-\lambda}a^{\lambda-1}}{\Gamma(1-\lambda)}\Gamma\left(\frac{1-\lambda+\mu+\nu}{2}\right)\Gamma\left(\frac{1-\lambda-\mu+\nu}{2}\right)\\
\times\Gamma\left(\frac{1-\lambda+\mu-\nu}{2}\right)\Gamma\left(\frac{1-\lambda-\mu-\nu}{2}\right) \, ,
\label{idenBesselK}
\end{multline}
which converges for $\operatorname{Re}a>0$, Re$\lambda<1-\vert\operatorname{Re}\mu\vert-\vert\operatorname{Re}\nu\vert$. The straightforward calculation shows that~\eqref{IdmSGII} holds for 
\begin{equation}
\mathfrak{C}_{n;\kappa}={2\kappa \choose n} \, , \quad \mathfrak{w}_{\kappa}(r)=\mathfrak{D}_{\kappa}\mathfrak{N}_{\kappa}(r) \, , \quad \mathfrak{D}_{\kappa}=\frac{4(2\kappa+1)}{[\Gamma(k+1)]^{2}} \, ,
\label{cnSGII}
\end{equation}
with ${a \choose b}$ the binomial coefficient. Thus, the expansion coefficients in the SGII CS define a binomial-like distribution weighted by a modified Bessel function of the second kind rather than the conventional binomial parameter.

Although the construction obtained so far for the SGII CS is general enough, henceforth we consider the particular case of half-integer values of $\kappa$, that is, $\kappa=L+1/2$ for $L\in\N$. In such a case, the Bessel function $K_{L+1/2}(z)$ writes as\cite{Mag54}
\begin{equation}
K_{L+\frac{1}{2}}(z)=\frac{e^{-z}}{\sqrt{2z}}\frac{\Gamma(L+1/2)}{(z/2)^{L}}\,{}_{1}F_{1}\left(\left.\begin{aligned} -L\,\,\\ -2L \, \end{aligned}\right\vert 2z\right) \, , \quad L=0,1,\cdots .
\label{KHG}
\end{equation}
We thus can exploit the symmetry of the modified Bessel functions of the second kind $K_{\nu}(z)=K_{-\nu}(z)$, together with the symmetry of the binomial coefficient ${2L+1 \choose n}$, in order to write the SGII CS as
\begin{equation}
\vert z;L\rangle_{\operatorname{II}}=\frac{1}{[\mathfrak{N}_{L}(r)]^{1/2}}\sum_{n=0}^{L}e^{\ii n \phi}c_{n;L}^{(\operatorname{II})}(r)\left[\vert n \rangle + e^{\ii(2L-2n+1)\phi} \vert 2L-n \rangle \right] \, ,
\label{CSCII}
\end{equation}
where the expansion coefficients are given by
\begin{equation}
c_{n;L}^{(\operatorname{II})}(r):=\sqrt{{2L+1\choose n}} \, \Gamma(L-n+1/2) r^{n} \, {}_{1}F_{1}\left(\left.\begin{aligned} n-L \, \,\\ 2n-2L \, \end{aligned}\right\vert\, 4r \right) \, ,
\label{cnCII}
\end{equation}
and the normalization simply reduces to
\begin{equation}
\mathfrak{N}_{L}(r)=2\sum_{n=0}^{L} \left[c_{n;L}^{(\operatorname{II})}(r)\right]^{2}.
\label{NCII}
\end{equation}
Given that $\mathfrak{N}_{L}(r)$ is a finite sum, and using the fact that $K_{n-L-1/2}(2r)$ is an entire function for $n=0,1,\cdots L$, we conclude that normalization function is well-defined on the whole complex-plane.

\section{Quantization map with SGII CS}
\label{subsec:QMCII}
Contrary to the coherent states introduced in Sec.~\ref{subsec:QmCI}, the quantization map defined through the family of SGII CS leads to linear mappings for the finite-dimensional vector space $\mathcal{H}^{(2\kappa)}$, for $\kappa\in\N^{*}/2$. In this form, any complex-valued function $j(z)$ induces a linear operator
\begin{equation}
j(z) \mapsto \mathbf{J}:=\int_{z\in\mathbb{C}}\frac{\ud^{2}z}{\pi}\mathfrak{w}_{\kappa}(r) j(z) \vert z;\kappa\rangle\langle z;\kappa\vert \, ,
\end{equation}
which can be thought as an endomorphism in $\mathcal{H}^{(2\kappa)}$, or a mapping operator of the form $\mathbf{J}:\mathcal{H}\rightarrow\mathcal{H}^{(2\kappa)}$. 
In particular, the functions $z$ and $\overline{z}$ lead to the annihilation and creation operators $z\mapsto\mathbf{c}^{(\kappa)}_{-}$ and $\overline{z}\mapsto \mathbf{c}^{(\kappa)}_{+}\equiv\left(\mathbf{c}^{(\kappa)}\right)^{\dagger}$, respectively, determined with the aid of~~\eqref{idenBesselK} as
\begin{align}
& \mathbf{c}^{(\kappa)}_{-}:= \sum_{n=0}^{2\kappa}\sqrt{(n+1)(2\kappa-n)}\vert n\rangle\langle n+1\vert \, , 
\label{asu11}\\
& \mathbf{c}^{(\kappa)}_{+}:= \sum_{n=0}^{2\kappa}\sqrt{(n+1)(2\kappa-n)}\vert n+1 \rangle\langle n\vert \, .
\label{ahsu11}
\end{align}
Notice that the commutation relation between the previous ladder operators give rise to a third operator of the form
\begin{equation}
-2\mathbf{c}_{0}:=[\mathbf{c}^{(\kappa)}_{-},\mathbf{c}^{(\kappa)}_{+}]=\sum_{n=0}^{2\kappa}(n-\kappa)\vert n\rangle\langle n\vert \, ,
\end{equation}
such that the set $\{\mathbf{c}_{-},\mathbf{c}_{+},\mathbf{c}_{0}\}$ represents a realization of the $\mathfrak{su}(2)$ Lie algebra,
\begin{equation}
[\mathbf{c}^{(\kappa)}_{0},\mathbf{c}^{(\kappa)}_{\pm}]=\pm\mathbf{c}^{(\kappa)}_{\pm} \, , \quad [\mathbf{c}^{(\kappa)}_{-},\mathbf{c}^{(\kappa)}_{+}]=-2\mathbf{c}^{(\kappa)}_{0} \, .
\end{equation}
It is interesting to notice that the operator $\mathbf{c}^{(\kappa)}_{0}$ plays an analogous role to that of the $z$-projection operator $\mathbf{L}_{z}$ of the total angular momentum $\overrightarrow{\mathbf{L}}$. The previous realization coincides with the \textit{Holstein-Primakoff} realization~\cite{Hol40,Nie97,Kap91} of SU$(2)$. 

Lastly, for the sake of completeness, it is worth mentioning that coherent states related to the set of generators $\{\mathbf{c}_{0}^{(\kappa)},\mathbf{c}_{-}^{(\kappa)},\mathbf{c}_{+}^{(\kappa)}\}$, constructed through the appropriate exponential representation, i.e, the so-called Radcliffe-Gilmore-Perelomov CS or spin CS, have been extensively discussed, see for instance~\cite{Per86,Gaz09,Kok10}. These $\mathfrak{su}(2)$ coherent states are  given by the finite linear combination
\begin{equation}
\vert \xi;\kappa\rangle_{\operatorname{\mathfrak{su}(2)}}=\sum_{n=0}^{2\kappa}\sqrt{{2\kappa \choose n}}\, \frac{\xi^{n}}{(1+\vert\xi\vert^{2})^{\kappa}}\vert n\rangle \, , \quad \xi\in\mathbb{C} \, ,
\label{su2CS}
\end{equation}
where $\kappa=1/2,1,3/2,2,\cdots$, as it occurs in the SGII CS.

\section{Boson realization and contraction of algebras}
\label{sec:Algebras}
In  previous works, see for instance ~\cite{Olm20} and references therein, it was found that the generators of the Fock representation of  $\mathfrak{su}(1,1)$  are achievable from the appropriately quantization map using the Perelomov $\mathfrak{su}(1,1)$ coherent states, also called Berezin-Toeplitz quantization. Analogous results have been found for the $\mathfrak{su}(2)$ algebra. Interestinlgy, the SU$(1,1)$ and SU$(2)$ coherent states can be reduced to the conventional Glauber-Sudarshan coherent states~\cite{Gla63}, defined through the linear combination
\begin{equation}
\vert \alpha \rangle_{GS} = \sum_{n=0}^{\infty} c_{n}^{SG}(\alpha)\vert n\rangle \, , \quad c_{n}^{SG}(\alpha)=e^{-\frac{\vert\alpha\vert^{2}}{2}}\frac{\alpha^{n}}{\sqrt{n!}} \, .
\label{GSCS}
\end{equation}
Such a reduction is achieved by means of the so-called \textit{contraction procedure}~\cite{Gaz09}, where the coherence parameter should be appropriately reparametrized while the Bargmann index tends to infinity.

On the other hand, from the matrix representation, it is possible to find a way to reduce the algebra generators to the boson operators. To this end, let us consider the generators given in Eqs.~\eqref{ah}-\eqref{ahd} and the limit procedure
\begin{equation}
\lim_{\kappa\rightarrow\infty}\frac{\mathbf{a}_{-}^{\kappa}}{\sqrt{2\kappa}} \rightarrow \mathbf{a} \, , \quad \lim_{\kappa\rightarrow\infty}\frac{\mathbf{a}_{+}^{\kappa}}{\sqrt{2\kappa}} \rightarrow \mathbf{a}^{\dagger} \, .
\end{equation}
Similarly, from the generators of the $\mathfrak{su}(2)$ algebra in Eqs.~\eqref{asu11}-\eqref{ahsu11} we get
\begin{equation}
\lim_{\kappa\rightarrow\infty}\frac{\mathbf{c}_{-}^{\kappa}}{\sqrt{2\kappa}} \rightarrow \mathbf{a} \, , \quad \lim_{\kappa\rightarrow\infty}\frac{\mathbf{c}_{+}^{\kappa}}{\sqrt{2\kappa}} \rightarrow \mathbf{a}^{\dagger} \, .
\end{equation}
Thus, in both cases the boson operators are recovered.

From the previous reduction of algebra, it is clear that the $\vert\alpha;\kappa\rangle_{\mbox{\tiny I}}$ and $\vert z;\kappa\rangle_{\mbox{\tiny II}}$ should reduce to the Glauber-Sudarshan coherent states~\eqref{GSCS} under the appropriate limit. For clarity, we consider each case separately.

\subsubsection*{Susskind-Glogower-I CS}
For this case, we consider the conventional contraction used for the SU$(1,1)$ CS, that is, we reparametrize the coherence parameter as $\vert\alpha\vert\equiv r=\sqrt{\frac{k}{2}}\vert z\vert$, where $\vert z\vert^{2}$ is the field intensity of the conventional coherent states, while the Bargmann index tends to infinity, $\kappa\rightarrow\infty$. We thus get the reparametrized expansion coefficients
\begin{multline}
c_{n;\kappa}^{(I)}\left(\sqrt{\frac{\kappa}{2}}\,\vert z\vert\right):=\frac{e^{\ii\phi n}\vert z\vert^{n}}{\sqrt{n!}} \left[{}_{1}F_{2}\left(\left.\begin{aligned} 1/2 \hspace{6mm}\, \\ \kappa+1,\kappa+1 \end{aligned}\right\vert\, -2\kappa \vert z\vert^{2}\right)\right]^{-1/2} \\
\times\frac{\sqrt{(2\kappa)_{n}}(k/2)^{n/2}}{(\kappa+1)_{n}}\sum_{q=0}^{\infty}\frac{\left(-\frac{k}{2}\vert z\vert^{2}\right)^{q}}{q!(n+\kappa+1)_{q}} \, .
\label{APPB1}
\end{multline}
Straightforward calculations show that, in the limit $\kappa\rightarrow\infty$, the generalized hypergeometric function ${}_{1}F_{2}(\cdot)$ converges to one. Simultaneously, the Pochhammer symbols simplify as $(2k)_{n}\rightarrow (2k)^{n}$, $(\kappa+1)_{n}\rightarrow\kappa^{n}$, and $(n+\kappa+1)_{1}\rightarrow\kappa^{q}$. The latter leads to the expansion coefficients of the Glauber-Sudarshan coherent states
\begin{equation}
\lim_{\kappa\rightarrow\infty}c_{n;\kappa}^{(I)}\left(\sqrt{\frac{\kappa}{2}}\,\vert z\vert\right) \rightarrow \frac{e^{\ii\phi n}\vert z\vert^{n}}{\sqrt{n!}}\sum_{q=0}^{\infty}\frac{\left(-\frac{1}{2}\vert z\vert^{2}\right)^{q}}{q!}=\frac{e^{-\frac{\vert z\vert^{2}}{2}}z^{n}}{\sqrt{n!}} \, , \quad z:=\vert z\vert e^{\ii\phi} \, .
\end{equation}

\subsubsection*{Suskind-Glogower-II CS}
The SGII CS~\eqref{CSCII} are such that the parameter $L=0,1,\cdots$ dictates the total number $2L +1$ of elements in the linear combination. Thus, to recover the Glauber-Sudarshan coherent states, it is necessary to consider the limit $L\rightarrow\infty$. Although the latter limit allows recovering the infinite linear combination we require, this is just one part of the contraction procedure. Now, let us consider the same reparametrization used in the SGI CS case, $\vert\alpha\vert\equiv r=\sqrt{\frac{L}{2}}\vert z\vert$, together with the limit $L\rightarrow\infty$. With the help of the Stirling approximation~\cite{Olv10} we obtain the expansion coefficients and the normalization constant
\begin{equation}
\lim_{L\rightarrow\infty}c_{n;L}^{(\operatorname{II})}\left(\frac{L}{2}\vert z\vert\right)\sim\sqrt{\frac{\pi}{2}}\left(\frac{L}{e}\right)^{L}\frac{\left(e^{i\phi}\vert z\vert\right)^{n}}{\sqrt{n!}} \, , \quad \lim_{L\rightarrow\infty}\mathfrak{N}_{L}\left(\frac{L}{2}\vert z\vert\right)\sim \frac{\pi}{2}\left(\frac{L}{e}\right)^{2L}e^{\vert z\vert^{2}} \, ,
\end{equation}
respectively. We thus obtain
\begin{equation}
\lim_{L\rightarrow\infty} \frac{c_{n;L}^{(\operatorname{II})}\left(\frac{L}{2}\vert z\vert\right)}{\left[\mathfrak{N}_{L}\left(\frac{L}{2}\vert z\vert\right)\right]^{1/2}}\sim e^{-\frac{1}{2}\vert z \vert^{2}}\frac{z^{n}}{\sqrt{n!}} \, , \quad z=e^{\ii\phi}\vert z\vert \, ,
\end{equation}
which is indeed the contraction to the Glauber-Sudarshan coherent states.

\section{Photon statistics and nonclassical properties of the SGI and SGII coherent states}
\label{sec:stats}
In this section, we explore the nonclassical properties related to the SGI and SGII CS. Notably, the nonclassicality associated with the conventional and modified Susskind-Glogower coherent states has been studied in~\cite{CurXX}. Since, for $\kappa=1$, the SGI CS reduce to the modified Susskind-Glogower CS, the nonclassical properties reduces to those studied in~\cite{CurXX}, and we are left to analyze such behavior for $\kappa>1$. Moreover, the SGII CS provide a finite linear combination that deserves special attention by itself as an alternative to the SU$(2)$ coherent states~\eqref{su2CS}.

Although we can perform several nonclassicality tests, we focus mainly on two, namely the photon distribution squeezing and the quadrature squeezing. The photon distribution squeezing is determined through the \textit{Mandel parameter}~\cite{Man79,Ros19c}, a quantity that dictates the deviation of the variance from the average number of photons. The Mandel parameter is determined through 
\begin{equation}
\mathcal{Q}:=\frac{(\Delta\mathbf{n})^{2}}{\langle\mathbf{n}\rangle}-1 \, , \quad (\Delta\mathbf{n})^{2}=\langle\mathbf{n}^{2}\rangle-\langle\mathbf{n}\rangle^{2} \, .
\label{Mandel}
\end{equation}
The vacuum state $\vert 0 \rangle$ and the Glauber-Sudarshan coherent states are special cases in which $\mathcal{Q}=0$, that is, the photon are distributed in the linear combination as a Poisson distribution. For the Fock states $\vert n\rangle$, with $n=1,2\cdots$, the Mandel parameter takes its lowest value $Q=-1$, and so the photon distribution is said to be maximally squeezed. Furthermore, photon distribution squeezing is achieved for $-1\leq \mathcal{Q}<0$. The related states are said to have sub-Poisson photon distribution, and are nonclassical. For $\mathcal{Q}>0$, the states are said to have a super-Poisson distribution, or equivalently, they are viewed as ``classical'' states.

On the other hand, squeezing can be explored in terms of the physical quadratures $\mathbf{x}=(\mathbf{a}+\mathbf{a}^{\dagger})/\sqrt{2}$ and $\mathbf{p}=-\ii(\mathbf{a}-\mathbf{a}^{\dagger})/\sqrt{2}$, which satisfy the commutation relation $[\mathbf{a},\mathbf{a}^{\dagger}]=\ii$. In the latter, the Planck's constant has been absorbed for simplicity. We thus have the Heisenberg uncertainty relation
\begin{equation}
(\Delta\mathbf{x})(\Delta\mathbf{p})\geq\frac{1}{2} \, , \quad (\Delta\mathbf{x})^{2}=\langle\mathbf{x}^{2}\rangle-\langle\mathbf{x}\rangle^{2} \, , \quad (\Delta\mathbf{p})^{2}=\langle\mathbf{p}^{2}\rangle-\langle\mathbf{p}\rangle^{2} \, .
\label{HUR}
\end{equation}
Then, we say that an arbitrary state $\vert\psi\rangle$ squeezes the quadrature $\mathbf{x}$ whenever the inequality $(\Delta\mathbf{x})^{2}<1/2$ holds~\cite{Kin01,Dod02}. Clearly, the uncertainty relation~\eqref{HUR} implies that the variance of the quadrature $\mathbf{p}$ spreads. Analogous results hold for $(\Delta \mathbf{p})^{2}<1/2$.

Now, we discuss the nonclassicality of the families of SGI and SGII coherent states separately.

\subsubsection*{Susskind-Glogower-I CS}
Let us consider the SGI CS, defined by the expansion coefficient given in~\eqref{CIcn}. We can exploit the same procedure used to determine the normalization constant $\mathcal{N}_{\kappa}(r)$ (for details see App.~\ref{sec:norm}) and compute the average value of the number of photons $\langle\mathbf{n}\rangle_{\kappa;I}\equiv\langle\alpha;\kappa\vert\mathbf{n}\vert\alpha;\kappa\rangle$, as well as $\langle\mathbf{n}^{2}\rangle_{\kappa;\operatorname{I}}\equiv\langle\alpha;\kappa\vert\mathbf{n}^{2}\vert\alpha;\kappa\rangle$. We thus get
\begin{align}
&\langle \mathbf{n} \rangle_{\kappa;\operatorname{I}} = \frac{\Gamma(2\kappa+1)r^{2}}{[\Gamma(\kappa+2)]^{2}\mathcal{N}_{\kappa}(r)}\, {}_{2}F_{3}\left(\left.\begin{aligned} 3/2,&1 \\ \kappa+2,&\kappa+2,2 \end{aligned}\right\vert -4r^{2} \right) = \frac{\Gamma(2\kappa+1)}{2[\Gamma(\kappa+1)]^{2}\mathcal{N}_{\kappa}(r)}-\kappa \, ,
\label{n1-CI}\\
&\langle \mathbf{n}^{2} \rangle_{\kappa;\operatorname{I}} = \langle \mathbf{n} \rangle_{\kappa} + \frac{\Gamma(2\kappa+2)r^{4}}{[\Gamma(\kappa+3)]^{2}\mathcal{N}_{\kappa}(r)}\, {}_{2}F_{3}\left(\left.\begin{aligned} 3/2,&2 \\ \kappa+3,&\kappa+3,3 \end{aligned}\right\vert -4r^{2} \right) \, .
\label{n2-CI}
\end{align}

\begin{figure}
\centering
\subfloat[][$\overline{n}_{\kappa}(r)$]{\includegraphics[width=0.3\textwidth]{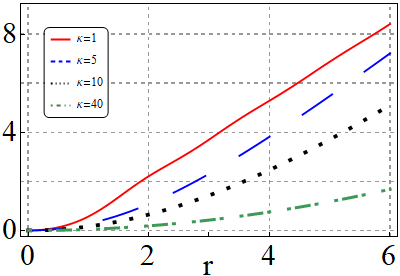}}
\hspace{1mm}
\subfloat[][$\frac{\ud\overline{n}_{\kappa}(r)}{\ud r}$]{\includegraphics[width=0.3\textwidth]{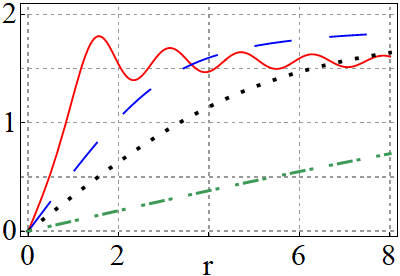}}
\hspace{1mm}
\subfloat[][$r(\overline{n})$]{\includegraphics[width=0.3\textwidth]{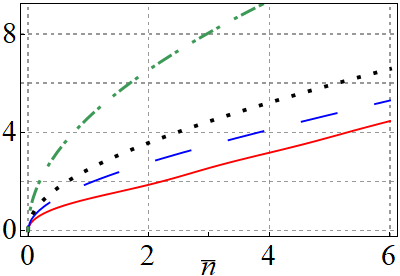}}
\caption{Average number of photons $\overline{n}_{\kappa}(r)\equiv\langle\mathbf{n}\rangle$ (a) and its respective derivative (b) as a function of the coherence parameter modulus $r\equiv\vert\alpha\vert$ for several values of $\kappa$. Coherence parameter modulus $r(\overline{n})$ as a function of the average number of photons (c).}
\label{Fig0}
\end{figure}

We have previously noticed a resemblance between the expansion coefficients of the SU$(1,1)$ CS and those of the SGI CS, as both are weighted by a negative binomial factor. To explore to what extend such states are indeed related, it is useful to analyze their respective photon distributions. Nevertheless, the SU$(1,1)$ CS~\eqref{Psu11} converge in the open unit-disk, whereas the SGI converge in the whole complex-plane. Therefore, it is necessary to establish common grounds in which both coherent states can be compared. Such a task is achieved by expressing every nonclassical measure or statistical properties in terms of the average number of photon $\overline{n}(\vert\alpha\vert)=\langle \alpha \vert\mathbf{n}\vert\alpha\rangle$, with $\vert \alpha\rangle$ any family of coherent states parametrized by a complex parameter $\alpha$. The latter is one of the global property that characterizes any quantum state of light constructed as a linear combination of Fock states. Henceforth, we rewrite $r=\vert\alpha\vert$ in terms of the average number of photons $\overline{n}$ for every coherent state in consideration, that is, $r(\overline{n})$. Although such reparametrization cannot always be achieved through analytic expressions, it is always feasible by numerical means. In this form, we can compare any two coherent states, even if those converge in different regions of the complex-plane. 

Before proceeding, it is worth to mention that, for the Glauber-Sudarshan case, the average number of photons is simply given by $\overline{n}(r)=r^{2}$, whereas for the SU$(1,1)$ CS we get $\overline{n}(r)=2\kappa r^{2}(1-r^{2})^{-1}$. Both functions are trivially invertible. For the SGI case, the average number of photons $\overline{n}_{\kappa;\operatorname{I}}(r)$ in~\eqref{n1-CI} is an invertible function if it is a strictly monotone function~\cite{Jef56,Roy88}, that is, for $r_{1}>r_{2}$ ($r_{1}<r_{2}$) we have $\overline{n}(r_{1})>\overline{n}(r_{2})$ ($\overline{n}(r_{1})<\overline{n}(r_{2})$). From~\eqref{n1-CI}, it is clear that $\overline{n}_{\kappa;\operatorname{I}}(r)>0$ for all $r\geq 0$. Moreover, it is trivial to prove that $\ud\overline{n}_{\kappa;\operatorname{I}}(r)/\ud r>0$, for all $r>0$. We conclude that $\overline{n}_{\kappa;\operatorname{I}}(r)$  is a strictly increasing function, and thus invertible for $r>0$. The behavior of the average number of photon and its numerical inverse function are depicted in Fig.~\ref{Fig0} for several values of $\kappa$.

\begin{figure}
\centering
\subfloat[][$n=1$]{\includegraphics[width=0.4\textwidth]{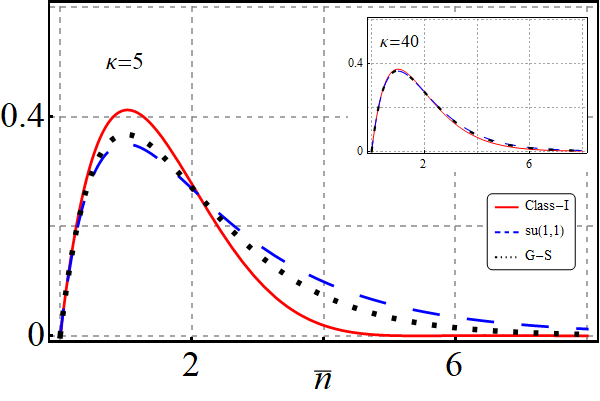}}
\hspace{5mm}
\subfloat[][$n=4$]{\includegraphics[width=0.4\textwidth]{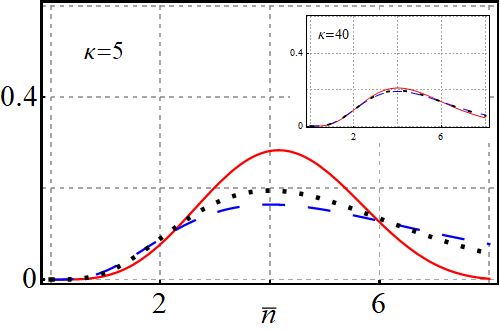}}
\caption{Probability densities $\mathcal{P}_{n}^{\operatorname{I}}(\overline{n};\kappa)$ (solid-red), $\mathcal{P}_{n}^{\operatorname{\mathfrak{su}(1,1)}}(\overline{n};\kappa)$ (dashed-blue), and $\mathcal{P}_{n}^{\operatorname{GS}}(\overline{n};\kappa)$ (dotted-black) in terms of their respective average number of photons $\overline{n}$ for $n=1$ (a) and $n=4$ (b). In both cases, $\kappa=5$ and $\kappa=40$ have been fixed for the main plot and the inset, respectively.} 
\label{Fig1}
\end{figure}

Now, we compare the photon statistics for several families of coherent states in terms of their respective average number of photons. To this end, let us consider the SU$(1,1)$, SGI, and Glauber-Sudarshan coherent states, together with the respective probability densities $\mathcal{P}_{n}^{\operatorname{\operatorname{SU}(1,1)}}(\overline{n};\kappa)=\vert c_{n;\kappa}^{\operatorname{\operatorname{SU}(1,1)}}(r(\overline{n}))\vert^{2}$, $\mathcal{P}_{n}^{\operatorname{I}}(\overline{n};\kappa)=\vert c_{n;\kappa}^{\operatorname{(I)}}(r(\overline{n}))\vert^{2}$, and $\mathcal{P}_{n}^{\operatorname{GS}}(\overline{n};\kappa)=\vert c_{n}^{\operatorname{GS}}(r(\overline{n}))\vert^{2}$. We thus depict the behavior of the probability densities in Fig.~\ref{Fig1} for $\kappa=5$ and $\kappa=40$ (see insets in Fig.~\ref{Fig1}). From the latter, it is clear that larger values of $\kappa$ steer both probability densities $\mathcal{P}_{n}^{\operatorname{\operatorname{SU}(1,1)}}(\overline{n};\kappa)$ and $\mathcal{P}_{n}^{\operatorname{I}}(\overline{n};\kappa)$ into $\mathcal{P}_{n}^{\operatorname{GS}}(\overline{n};\kappa)$. Clearly, such a behavior corresponds to the contraction limit of the $\mathfrak{su}(1,1)$ coherent states~\cite{Gaz09} and to the contraction limit discussed in Sec.~\ref{sec:Algebras}.

\begin{figure}[H]
\centering
\subfloat[][$\mathcal{Q}$]{\includegraphics[width=0.42\textwidth]{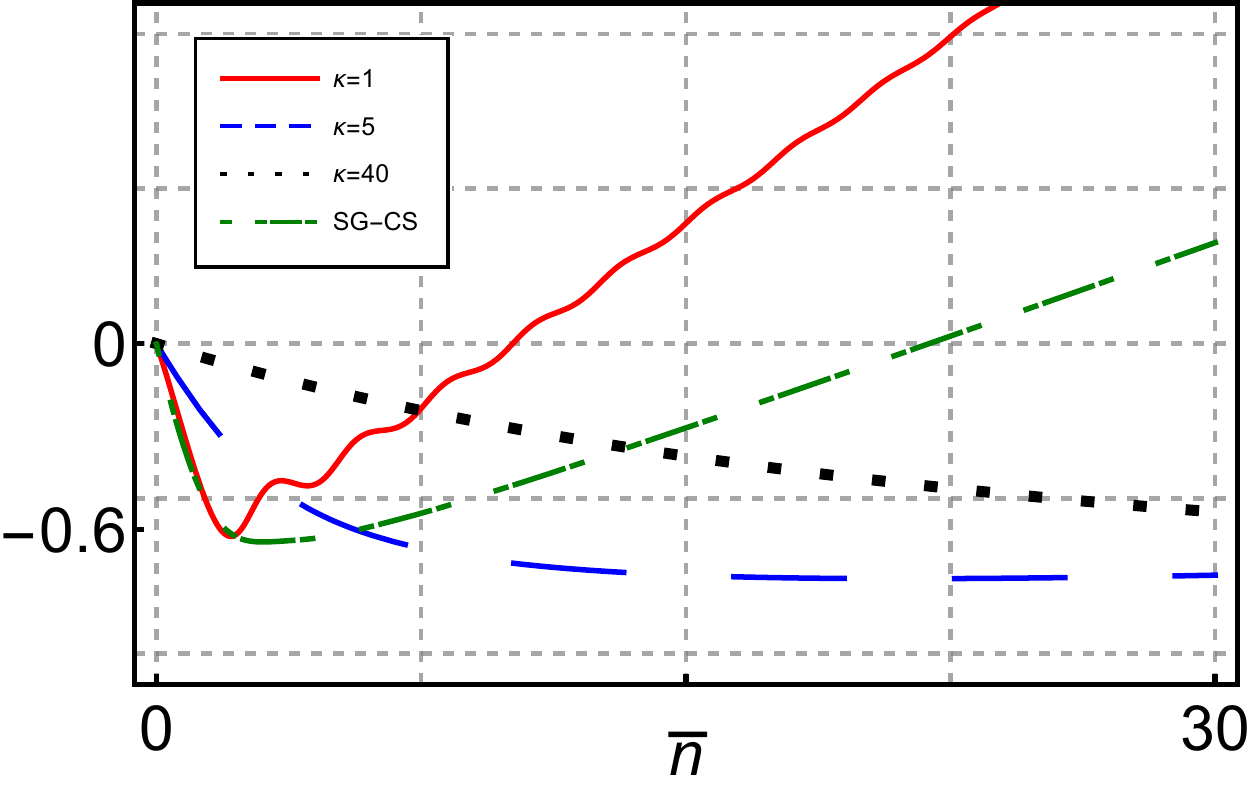}
\label{ManCI}}
\hspace{2mm}
\subfloat[][$(\Delta \mathbf{x})^{2}$]{\includegraphics[width=0.4\textwidth]{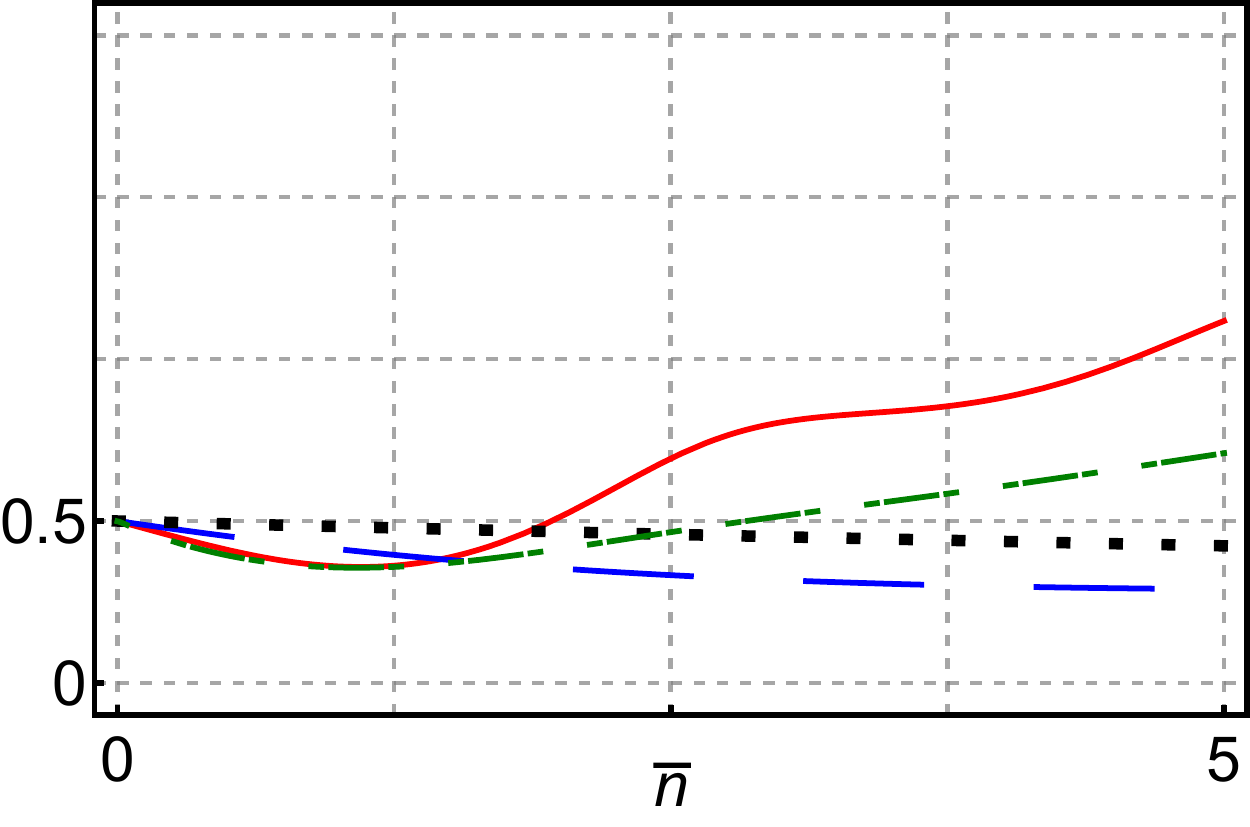}
\label{DXCI}}
\hspace{2mm}
\subfloat[][$(\Delta \mathbf{p})^{2}$]{\includegraphics[width=0.4\textwidth]{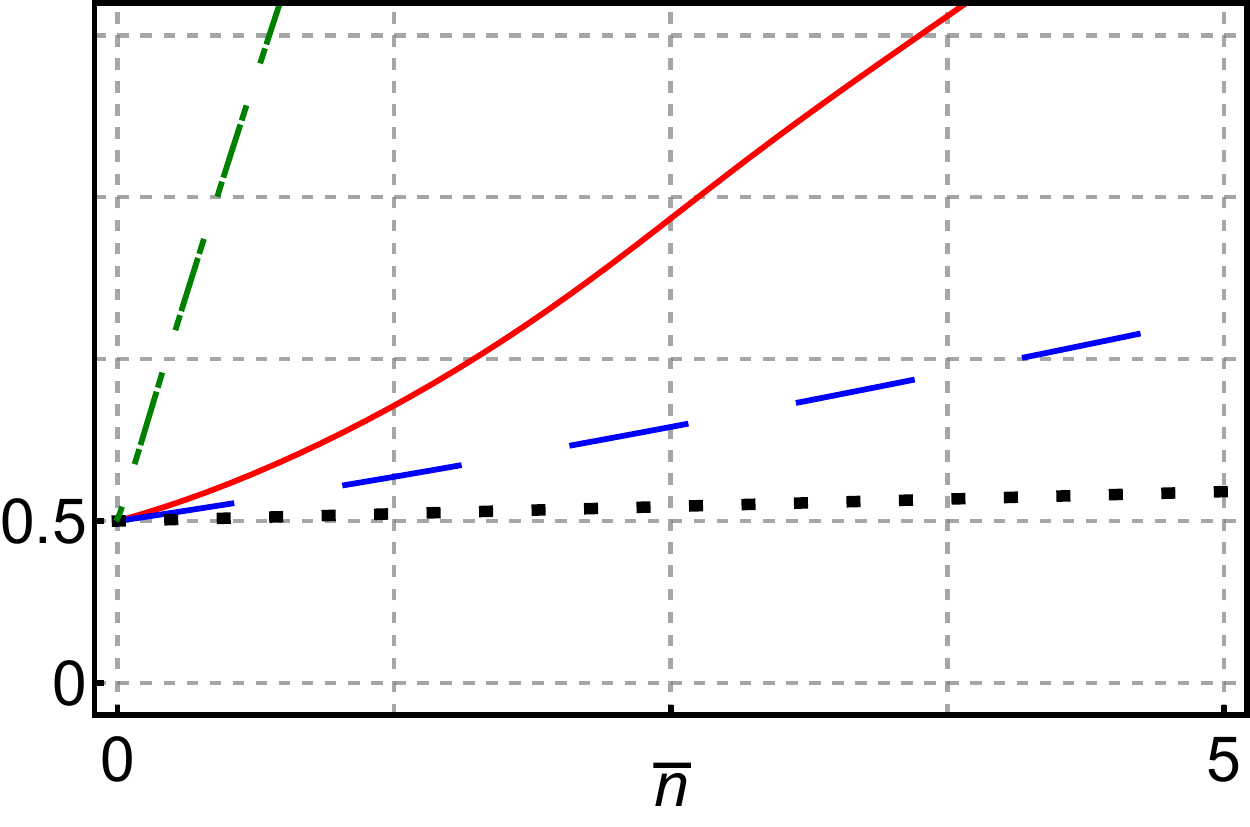}
\label{DPCI}}
\caption{Mandel parameter~\eqref{Mandel} (a), together with the physical variances $(\Delta \mathbf{x})^{2}$ (b) and $(\Delta\mathbf{p})^{2}$ (c) for the SGI CS~\eqref{opSGcs} as a function of the average number of photons~\eqref{n1-CI} for several values of $\kappa$.} 
\label{Fig2}
\end{figure}

It is worth recalling that the SU$(1,1)$ coherent states emerge naturally in the construction of squeezed states of light, an archetype example of a nonclassical state of light~\cite{Wod85,Kok10}. Thus, given the relation between the SGI and the SU$(1,1)$ coherent states, it is expected to obtain some interesting nonclassical properties from the SGI CS. Furthermore, for the time being, we are mostly interested to compare the nonclassical properties among $\vert\alpha\rangle_{\mbox{\tiny SG}}$, $\vert\alpha\rangle_{\mbox{\tiny mSG}}$, and $\vert \alpha;\kappa\rangle_{\mbox{\tiny I}}$. Such an information is available in Fig.~\ref{ManCI}, where we depict the behavior for the Mandel parameter as a function of $\overline{n}$, with $\overline{n}$ the respective average number of photons for each family of coherent states under consideration. 

Although both the conventional and modified Susskind-Glogower ($\kappa=1$) coherent states have a negative Mandel parameter (photon distribution squeezing), it is clear that the former is more squeezed than the latter, a result already discussed in~\cite{CurXX}. Nevertheless, for higher values of $\kappa$, the SGI CS are indeed a more reliable source of nonclassicality, for the Mandel parameter becomes negative for larger values of the average number of photons. For completeness, in Figs.~\ref{DXCI}-\ref{DPCI} we depict the quadrature variances $(\Delta \mathbf{x})$ and $(\Delta \mathbf{p})$, where it follows that the SGI CS serve as a source of field-amplitude noise squeezing as well. However, the latter is highly sensitive to any change in the average number of photons. 

\subsubsection*{Susskind-Glogower-II CS}
\begin{figure}
\centering
\subfloat[][$\kappa=2$]{\includegraphics[width=0.4\textwidth]{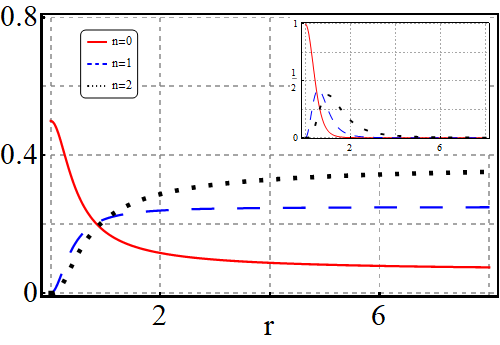}}
\hspace{5mm}
\subfloat[][$\kappa=5/2$]{\includegraphics[width=0.4\textwidth]{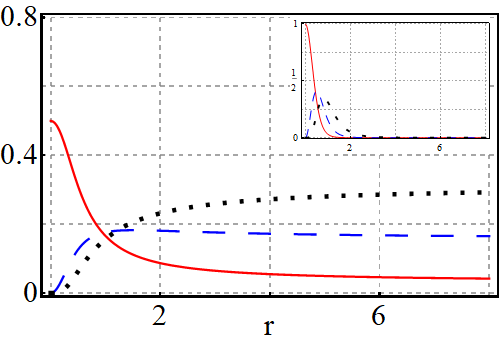}}
\caption{Probability densities $\mathcal{P}_{n}^{\operatorname{II}}(r;\kappa)$ in terms of $r=\vert\alpha\vert$ for $n=0$ (solid-red),  $n=1$ (dashed-blue), and $n=2$ (dotted-black). In the inset, we depict $\mathcal{P}_{n}^{\operatorname{\mbox{\tiny SU}(2)}}(r;\kappa)$ with the same parameters used in the main figure.} 
\label{ProbCII}
\end{figure}

Contrary to the SGI CS, the SGII CS are constructed as a finite linear combination of Fock states, and the average number of photons $\overline{\mathbf{n}}(r)$ is a bounded function because it cannot exceed the maximum number of photons used in the combination. From the symmetry fulfilled by the modified Bessel functions of the second kind, $K_{-\nu}(z)=K_{\nu}(z)$, and after some calculations, we find that the average number of photons is independent of $r$, $\overline{n}_{\kappa;\operatorname{II}}(r)=\kappa$. Thus, a reparametrization for $r=\vert z\vert$ in terms of the number of photons is not longer feasible. Despite this issue, we can compare the probability distribution for the SGII CS, $\mathcal{P}_{n}^{\mbox{\tiny II}}(r;\kappa)=\vert c_{n;\kappa}^{\mbox{\tiny (II)}}(r)\vert^{2}$, with that of the SU$(2)$ CS, $\mathcal{P}_{n}^{\operatorname{\mbox{\tiny SU}(2)}}(r;\kappa)=\vert c_{n;\kappa}^{\mbox{\tiny SU}(2)}(r)\vert^{2}$. The respective behavior is depicted in Fig.~\ref{ProbCII} for $\kappa=2$ and $\kappa=5/2$ for several values of $n$. In such a figure, the probability distribution for the SGII CS behaves as a constant, different from zero, for large values of $r$. The latter is a result that can be determined explicitly from the asymptotic behavior, $K_{\nu}(x)\sim \sqrt{\frac{\pi}{2x}}e^{-x}$, for $x>>1$. In this form, we find 
\begin{equation}
C_{n;\kappa}^{\operatorname{(II)}}(r)\sim \frac{e^{in\phi}}{2^{k}}\sqrt{{2\kappa \choose n}} \, , \quad r>>1 \, , \quad z=re^{i\phi} \, ,
\label{CIIAsym}
\end{equation}
which also explains the behavior depicted in Fig.~\ref{ProbCII} for large values of $r$.

\begin{figure}
\centering
\includegraphics[width=0.5\textwidth]{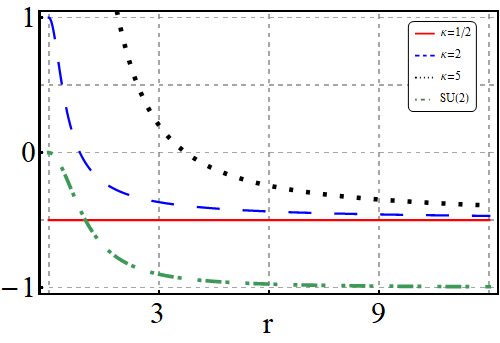}
\caption{Mandel parameter $\mathcal{Q}$, computed from~\eqref{Mandel} using the SGII CS $\vert z;\kappa\rangle_{\operatorname{II}}$, as a function of $r=\vert z\vert$ for $\kappa=1/2$ (solid-red), $\kappa=2$ (dashed-blue), and $\kappa=5$ (dotted-black). The dash-dotted-green curve denotes the Mandel parameter for the SU$(2)$ CS, which is independent of $\kappa$.}
\label{MandelCII}
\end{figure}

Although we do not have a simple expression for $\langle\mathbf{n}^{2}\rangle_{\kappa;\operatorname{II}}(r):={}_{\operatorname{II}}\langle z;\kappa\vert \mathbf{n}^{2}\vert z;\kappa\rangle_{\operatorname{II}}$, we can exploit the asymptotic coefficients~\eqref{CIIAsym}, leading to $\langle\mathbf{n}^{2}\rangle_{\kappa;\operatorname{II}}(r)\sim \kappa(\kappa+1/2)$ and to the Mandel parameter $\mathcal{Q}_{\operatorname{II}}\sim-1/2$. Notice that $\mathcal{Q}$ is independent of $\kappa$ and $r$ for $r>>1$. The global behavior of the Mandel parameter is depicted in Fig.~\ref{MandelCII}, where we compare the SGII and the SU$(2)$ coherent states. In it, we can see a photon-distribution transitioning from the super-Poisson to the sub-Poisson regimes as $r$ increases and approaching asymptotically to $\mathcal{Q}=-1/2$. Therefore, the SGII CS are indeed a source of nonclassical states; however, its photon distribution squeezing is lower than that of the SU$(2)$ CS.

\begin{figure}[H]
\centering
\subfloat[][$\kappa=1/2$]{\includegraphics[width=0.4\textwidth]{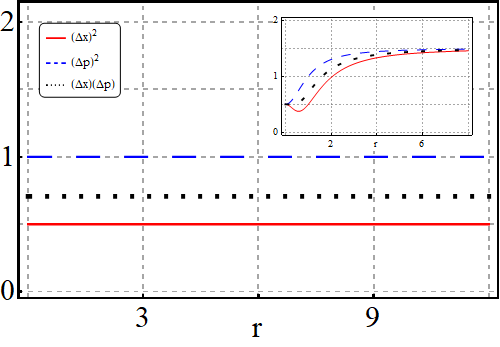}
\label{QuadCIIa}}
\hspace{2mm}
\subfloat[][$\kappa=2$]{\includegraphics[width=0.4\textwidth]{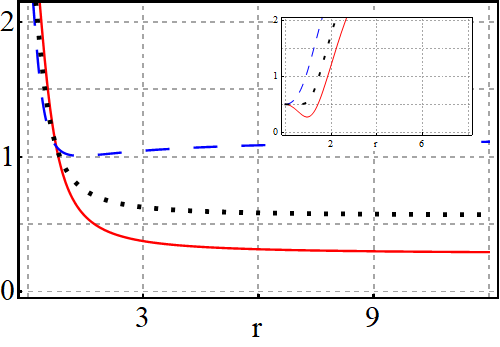}
\label{QuadCIIb}}
\hspace{2mm}
\subfloat[][$\kappa=5$]{\includegraphics[width=0.4\textwidth]{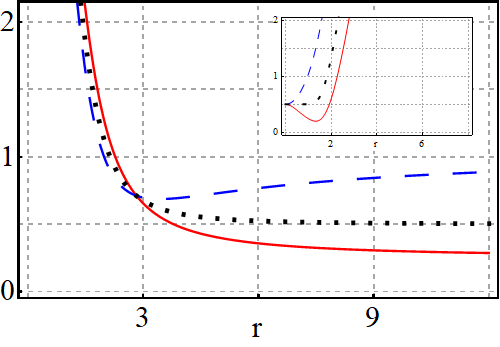}
\label{QuadCIIc}}
\caption{Quadrature variances $(\Delta \mathbf{x})^{2}$ (solid-red), $(\Delta \mathbf{p})^{2}$ (dashed-blue), and the product $(\Delta \mathbf{x})(\Delta \mathbf{p})$ (dotted-black) associated with the SGII CS. The inset denotes the respective variances for the SU$(2)$ coherent states.}
\label{QuadCII}
\end{figure}

On the other hand, the quadrature squeezing reveals complementary information to that of the photon distribution, which is available in Fig.~\ref{QuadCII}. In it, we compare the quadrature squeezing of both the SGII and the SU$(2)$ coherent states for several values of $\kappa$. Particularly, quadrature squeezing is missing for the SGII CS when $\kappa=1/2$, as shown in Fig.~\ref{QuadCIIa}. Nevertheless, for higher values of $\kappa$, we indeed obtain quadrature squeezing for the SGII CS, as shown in Figs.~\ref{QuadCIIb}-\ref{QuadCIIc}. In contradistinction to the SU$(2)$ CS, the quadrature squeezing distributes uniformly at large enough values of $r$ for the SGII CS, converging asymptotically to a constant value. This means that quadrature squeezing is preserved continuously for the SGII CS, where in the SU$(2)$ CS such a squeezing is highly susceptible to any changes on $r$. The latter clearly represents an advantage of the SGII CS over the SU$(2)$. 

\section{Conclusions}
\label{sec:Conclu}
Throughout this manuscript, we have successfully generalized the family of modified Susskind-Glogower coherent states~\cite{CurXX}, initially introduced as a deformation of the conventional Susskind-Glogower~\cite{Mon11a,Mon11b}. The latter was achieved by noticing that the quantization map of the appropriate optical functions associated with the modified Susskind-Glogower CS leads to generators of Fock representations of the $\mathfrak{su}(1,1)$ algebra, in which the Fock space plays the role of the unitary irreducible representation with the lowest Bargmann index, $\kappa=1$. The latter certainly paves the way for a generalization of the modified family so that the Bargmann index can be fixed freely. Such a task is feasible after cleverly modifying the index of the Bessel functions of the first kind inside the expansion coefficients. In this form, the resulting states, called Susskind-Glogower-I coherent states, are determined so that they satisfy both the completeness and normalization condition. In this form, we get the expected result, that is, the Fock basis carries the irreducible representation of $\mathfrak{su}(1,1)$ for arbitrary values of the Bargmann index $\kappa\geq 1$. 

After replacing the expansion coefficients of the modified Susskind-Glogower CS with modified Bessel function of the second kind, while truncating the linear combination to avoid singularities, we managed to construct the normalized Suskind-Glogower-II CS, which are a family of overcomplete states in a Hilbert space of finite-dimension. This new family of states is such that, in contradistinction to the SGI CS, we obtain the generators of $\mathfrak{su}(2)$ after performing the appropriate quantization map. Interestingly, using the contraction limit~\cite{Gaz09}, both the SGI and the SGII CS reduce to the conventional Glauber-Sudarshan CS, while the respective algebra generators reduce to the boson operators. That is, the algebras $\mathfrak{su}(1,1)$ and $\mathfrak{su}(2)$  contract to the Heisenberg-Weyl algebra. 

Remarkably, the new families of coherent states exhibit nonclassical properties that enhance some of the properties for the SU$(1,1)$ and SU$(2)$ coherent states. It is worth to remark that, for the SGII CS, we obtain quadrature and photon squeezing that stabilizes for $r\gg1$, that is, the nonclassical properties converge smoothly to a constant value, as it was depicted in Figs.~\ref{ProbCII}-\ref{QuadCII}. Particularly, although the Mandel parameter related to the SGII CS is less negative (less nonclasssical) than that of the SU$(2)$ CS, the quadrature squeezing is larger and constant for the SGII CS.

The results obtained so far pave the way for further research. For instance, after comparing the SGI CS with the SU$(1,1)$ coherent states, it is clear that both share the same negative-binomial factor weighted by different functions of the coherence parameter modulus $r=\vert\alpha\vert$. Thus, given that the resulting quantization map generates the same Lie algebra, it is clear that a relationship between the two families of coherent states should exist. Since both families of coherent states are normalized, there must be a unitary transformation that connects them. The latter could bring a way to determine the appropriate displacement operator to construct the SGI CS. Analogous results should hold for the SGII CS. On the other hand, the results obtained so far have led to the one-photon realization of the $\mathfrak{su}(1,1)$ and $\mathfrak{su}(2)$ algebras. However, it is still possible to explore to what extend multiboson realizations are feasible, such as the well-known two-photon realization introduced for the squeezed states by Nieto et al.~\cite{Nie97}. The latter requires special attention by itself and the corresponding results will be reported elsewhere.


\appendix

\section{Normalization constant $\mathcal{N}_{\kappa}(r)$}
\label{sec:norm}
In this section, we show the intermediate step required to compute the normalization constant associated to the SGI CS. To this end, we exploit the series expansion of the Bessel function $J_{\nu}(z)$ given in~\eqref{besselJ}. We thus get
\begin{equation}
\mathcal{N}_{\kappa}(r)=\sum_{n=0}^{\infty}\frac{(n+1)_{2\kappa-1}}{r^{2\kappa}}[J_{n+\kappa}(2r)]^{2}=\sum_{n,p,q=0}^{\infty}\frac{(-1)^{p+q}(n+1)_{2\kappa-1}r^{2(n+p+q)}}{p!q!\Gamma(p+n+k+1)\Gamma(q+n+k+1)} \, .
\end{equation}
The latter can be taken into a convenient form after noticing that terms inside a double-sum can be re-arranged as
\begin{equation}
\sum_{n=0}^{\infty}\sum_{m=0}^{\infty}A_{n,m}=\sum_{n=0}^{\infty}\sum_{m=0}^{n}A_{n-m,m}=\sum_{m=0}^{\infty}\sum_{n=0}^{m}A_{n,m-n} \, ,
\end{equation}
where, after performing two times the previous identity, we have
\begin{equation}
\mathcal{N}_{\kappa}(r)=\sum_{n=0}^{\infty}\frac{r^{2n}}{[(n+\kappa)!]^{2}}\sum_{p=0}^{n}\sum_{q=0}^{n-p}\frac{(-1)^{p+q}(n+1-p-q)_{2\kappa-1}}{p!q!(n+\kappa+1)_{-q}(n+\kappa+1)_{-p}} \, .
\end{equation}
Interestingly, the finite sum on $q$ can be re-arranged once again after applying the change of variable $q\rightarrow n-p-q$, leading to
\begin{equation}
\mathcal{N}_{\kappa}(r)=\sum_{n=0}^{\infty}\frac{(-1)^{n}r^{2n}}{[(n+\kappa)!]^{2}}\sum_{p=0}^{n}\frac{1}{p!(n+\kappa+1)_{-p}}\sum_{q=0}^{n-p}\frac{(-1)^{q}(q+1)_{2\kappa-1}}{(n-p-q)!(n+\kappa+1)_{q+p-n}} \, .
\label{Nnorm3}
\end{equation}
The latter allows reducing the sum over $q$ into a hypergeometric function of the form
\begin{equation}
\frac{\Gamma(2\kappa)\Gamma(n+\kappa+1)}{\Gamma(n+1-p)\Gamma(p+\kappa+1)}{}_{2}F_{1}\left(\left.\begin{aligned} 2\kappa, p-n \\ p+\kappa+1\end{aligned}\, \right\vert 1 \right)=\frac{\Gamma(2\kappa)\Gamma(n+1-\kappa)}{\Gamma(n+1-p)\Gamma(p+1-\kappa)}
\end{equation}
where the asymptotic behavior ${}_{2}F_{1}\left(\left.\begin{aligned} a&,b \\ &c \end{aligned}\right\vert 1\right)=\frac{\Gamma(c)\Gamma(c-a-b)}{\Gamma(c-a)\Gamma(c-b)}$ have been used~\cite{Olv10}. After substituting the previous result in ~\eqref{Nnorm3} we reduced the normalization constant to the double sum
\begin{equation}
\mathcal{N}_{\kappa}(r)=\Gamma(2\kappa)\sum_{n=0}^{\infty}\frac{r^{2n}(-1)^{n}(1-\kappa)_{n}}{[\Gamma(n+\kappa+1)]^{2}n!}\sum_{p=0}^{n}\frac{(-n-\kappa)_{p}(-n)_{p}}{p!(1-\kappa)_{p}} \, .
\end{equation}
The sum over $p$ is treated in similar form to that of $q$. After some calculations we get the final result
\begin{equation}
\mathcal{N}_{\kappa}(r)=\frac{\Gamma(2\kappa)}{[\Gamma(k+1)]^{2}}\, \, {}_{1}F_{2}\left(\left.\begin{aligned} &1/2 \\ \kappa+1&,\kappa+1 \end{aligned}\right\vert -4r^{2} \right) \, .
\end{equation}


\subsection*{Acknowledgments}
J.-P. Gazeau acknowledges partial support of CNRS-CRM-UMI 3457. V. Hussin acknowledges the support of research grants from NSERC of Canada. J. Moran acknowledges the support of the D\'epartement de Physique at the Universit\'e de Montr\'eal. K. Zelaya acknowledges the support from the Mathematical Physics Laboratory of the Centre de Recherches Math\'ematiques, through a postdoctoral fellowship. K. Zelaya also acknowledges the support of Consejo Nacional de Ciencia y Tecnolog\'ia (Mexico), grant number A1-S-24569. 




\begin{thebibliography}{99}

\bibitem{Sch26}
E. Schr\"odinger, \textit{Collected Papers on Wave Mechanics} 3rd (Augmented) English Edn, Chelsea Publishing, Providence, 1982.

\bibitem{Gla63}
R.J. Glauber, The Quantum Theory of Optical Coherence, \textit{Phys. Rev.} \textbf{130} (1963) 2529; R.J. Glauber, Photon Correlations, \textit{Phys. Rev. Lett.} \textbf{10} (1963) 84; R.J. Glauber, Coherent and incoherent states of the radiation field, \textit{Phys. Rev. Lett.} \textbf{10} (1963) 84.

\bibitem{Sud63}
E.C.G. Sudarshan, Equivalence of semiclassical and quantum mechanical descriptions of statistical light beams, \textit{Phys. Rev. Lett.} \textbf{10} (1963) 277.

\bibitem{Man65}
L. Mandel and E. Wolf, Coherence Properties of Optical Fields, \textit{Rev. Mod. Phys.} \textbf{37} (1965) 231.

\bibitem{Man79}
L. Mandel, Sub-Poissonian photon statistics in resonance fluorescence, \textit{Opt. Express} \textbf{4} (1979) 205.

\bibitem{Gaz09}
J.-P. Gazeau, \textit{Coherent States in Quantum Physics}, Wiley-VCH, Germany, 2009.

\bibitem{Kla00}
J.R. Klauder, \textit{Beyond Conventional Quantization}, Cambridge University Press, Cambridge, 2000.

\bibitem{Per04}
A. Peres and D.R. Tern, Quantum information and relativity theory, \textit{Rev. Mod. Phys.} \textbf{76} (2004) 93.

\bibitem{Gaz15}
J.-P. Gazeau and B. Heller, Positive-Operator Valued Measure (POVM) Quantization, \textit{Axioms} \textbf{4} (2015) 1.

\bibitem{Kim02}
M.S. Kim, W. Son, V. Bu\v{z}ek, and P.L. Knight, Entanglement by a beam splitter: Nonclassicality as a prerequisite for entanglement, \textit{Phys. Rev. A} \textbf{65} (2002) 032323.

\bibitem{Rad71}
J.M. Radcliffe, Some properties of coherent spin states, \textit{J. Phys. A: Gen. Phys.} \textbf{4} (1971) 313.

\bibitem{Are72}
F.T. Arecchi, E. Courtens, R. Gilmore, and H. Thomas, Atomic Coherent States in Quantum Optics, \textit{Phys. Rev. A} \textbf{6} (1972) 2211.

\bibitem{Ros16}
O. Rosas-Ortiz, S. Cruz y Cruz, M. Enr\'iquez, SU$(1,1)$ and SU$(2)$ approaches to the radial oscillator: Generalized coherent states and squeezing variances, \textit{Ann. Phys.} \textbf{373} (2016) 346.

\bibitem{Per86}
A. Perelomov, \textit{Generalized Coherent States and Their Applications}, Springer-Verlag, Berlin, 1986.

\bibitem{Wod85}
K. W\'odkiewicz, and J.H. Eberly, Coherent states, squeezed fluctuations, and the SU$(2)$ and SU$(1,1)$ groups in quantum-optics applications, \textit{J. Opt. Soc. Am. B} \textbf{2} (1985) 458.

\bibitem{Ber93}
Y. Berube-Lauziere, V. Hussin, Comments of the definitions of coherent states for the SUSY harmonic oscillator, \textit{J. Phys. A: Math. Gen.} \textbf{26} (1993) 6271 .

\bibitem{Mat96}
R.L. de Matos Filho, W. Vogel, Nonlinear coherent states, \textit{Phys. Rev. A} \textbf{54} (1996) 4560.

\bibitem{Man97}
V.I. Man'ko, G. Marmo, E.C.G. Sudarshan and F. Zaccaria, f-Oscillators and Nonlinear Coherent States, \textit{Phys. Scr.} \textbf{55} (1997) 528.

\bibitem{Jun99}
G. Junker, P. Roy, Non-linear coherent states associated with conditionally exactly solvable problems, \textit{Phys. Lett. A} \textbf{257} (1999) 113.

\bibitem{Zel17}
K. Zelaya, O. Rosas-Ortiz, Z. Blanco-Garc\'ia, and S. Cruz y Cruz, Completeness and Nonclassicality of Coherent States for Generalized Oscillator Algebras, \textit{Adv. Math. Phys.} \textbf{2017} (2017) 7168592.

\bibitem{Moj18}
B. Mojarevi, A. Dehghani, R.J. Bahrbeig, Excitation on the para-Bose states: nonclassical properties, \textit{Eur. Phys. J. Plus} \textbf{133} (2018) 34.

\bibitem{Wal83}
D.F. Walls, Squeezed states of light, \textit{Nature} \textbf{306} (1983) 141.

\bibitem{Lou87}
R. Loudon and P. L. Knight, Squeezed light, \textit{J. Mod. Opt.} \textbf{34} (1987) 709.

\bibitem{Tei89}
M.C. Teich and B.E.A. Saleh, Squeezed state of light, \textit{Quantum Opt.: J. Euro. Opt. Soc. Part B} \textbf{1} (1989) 153 .

\bibitem{Zel18}
K. Zelaya, S. Dey, and V. Hussin, Generalized squeezed states, \textit{Phys. Lett. A} \textbf{382} (2018) 3369.

\bibitem{Zel18A}
O. Rosas-Ortiz, and K. Zelaya, Bi-orthogonal approach to non-Hermitian Hamiltonians with the oscillator spectrum: Generalized coherent states for nonlinear algebras, \textit{Ann. Phys.} \textbf{388} (2018) 26.

\bibitem{Dey18}
S. Dey, A. Fring, and V. Hussin, A Squeezed Review on Coherent States and Nonclassicality for Non-Hermitian Systems with Minimal Length. In: \textit{Antoine JP., Bagarello F., Gazeau JP. (eds) Coherent States and Their Applications}, Springer Proceedings in Physics, vol \textbf{205}, Cham, 2018.

\bibitem{Gue18}
J. Guerrero, Non-Hermitian Coherent States for Finite-Dimensional Systems. In: Antoine JP., Bagarello F., Gazeau JP. (eds) Coherent States and Their Applications. Springer Proceedings in Physics, vol \textbf{205}. Springer, Cham, 2018.

\bibitem{Bag20}
F. Bagarello, Susy for Non-Hermitian Hamiltonians, with a View to Coherent States, \textit{Phys. Anal. Geom.} \textbf{23} (2020) 28.

\bibitem{Rec08}
J. R\'ecamier, M. Gorayeb, W. L. Moch\'an, and J. L. Paz, Nonlinear Coherent States and Some of Their Properties, \textit{Int. J . Theor. Phys.} \textbf{47} (2008) 673.

\bibitem{Mon11a}
R. De J. Le\'on Montel and H. M. Moya-Cessa, Modeling non-linear coherent states in fiber arrays, \textit{Int. J. Quant. Information} \textbf{9} (2011) 349.

\bibitem{Mon11b}
R. De J. Le\'on Montel, H. M. Moya-Cessa and F. Soto-Eguibar, Nonlinear coherent states for the Susskind-Glogower operators, \textit{Rev. Mex. F\'is.} \textbf{57} (2011) 133.

\bibitem{Gaz19}
J.-P. Gazeau, Coherent States in Quantum Optics: An Oriented Overview, \textit{in Integrability, Supersymmetry and Coherent States: A Volume in Honour of Professor Véronique Hussin}; \c{S}eng\"ul Kuru, Javier Negro, and Luis M. Nieto (Eds), Springer, 2019.

\bibitem{CurXX}
E.M.F. Curado, S. Faci, J.-P. Gazeau, and D. Nogera, Lowering Helstrom bound with non-standard coherent states, arXiv:2010.00171 [quant-ph].

\bibitem{Sus64}
L. Susskind, and J. Glogower, Quantum mechanical phase and time operator, \textit{Phys.} \textbf{1} (1964) 49.

\bibitem{Dir35}
P.A.M. Dirac, \textit{The Principles of Quantum Mechanics}; 2nd. edn., Claredon, Oxford, 1935.

\bibitem{Nik88}
A.F. Nikiforov and V.B. Uvarov, \textit{Special Functions of Mathematical Physics: A Unified Introduction with Applications}, Birkh\"auser, Germany,1988.

\bibitem{AliXX} S.T. Ali, J.P. Antoine, and J.P. J.P.Gazeau, Continuous Frames in Hilbert Space, \textit{Annals of Physics}
 \textbf{222} (1993) 1-37.

\bibitem{Gos16}
M.A. de Gosson, \textit{Born-Jordan Quantization: Theory and Applications}, Springer International Publishing, Switzerland, 2016. 

\bibitem{Mag54}
W. Magnus, and F. Oberhettinger, \textit{Formulas and Theorems for the Functions of Mathematical Physics}, Chelsea Publishing, New York, 1954.

\bibitem{Olm20}
M.A. del Olmo, and J.-P. Gazeau, Covariant integral quantization of the unit disk, \textit{J. Math. Phys.} \textbf{61} (2020) 022101.

\bibitem{Gra07}
I.S. Gradshteyn, and I.M. Ryzhik, \textit{Table of Integrals, Series, and Products, 7th ed.}, Academic Press, California, 2007.

\bibitem{Olv10}
F.W.J. Oliver, \textit{et al.} (eds.), \textit{NIST Handbook of Mathematical Functions}, Cambridge University Press, New York, 2010.

\bibitem{Nie97}
M.M. Nieto and D.R. Truax, Holstein-Primakoff/Bogoliubov transformations and the multiboson system, \textit{Fortsch.Phys.} \textbf{45} (1997) 145.

\bibitem{Ger05}
C.C. Gerry, and P.L. Knight, \textit{Introductory Quantum Optics}, Cambridge University Press, Cambridge, 2005.

\bibitem{Kok10}
P. Kok, and B.W. Lovett, \textit{Introduction to Optical Quantum Information Processing}, Cambridge University Press, Cambridge, 2010.

\bibitem{Hol40}
T. Holstein and H. Primakoff, Field Dependence of the Intrinsic Domain Magnetization of a Ferromagnet, \textit{Phys. Rev.} \textbf{58} (1940) 1098.

\bibitem{Kap91}
D.V. Kapor, M.J. \v{S}krinjar, and S.D. Stojanovi\'c, Relation between spin-coherent states and boson-coherent states in the theory of magnetism, \textit{Phys. Rev. B} \textbf{44} (1991) 2227.

\bibitem{Ros19c}
O. Rosas-Ortiz, Coherent and Squeezed States: Introductory Review of Basic Notions, Properties, and Generalizations, \textit{in Integrability, Supersymmetry and Coherent States: A Volume in Honour of Professor Véronique Hussin}; \c{S}eng\"ul Kuru, Javier Negro, and Luis M. Nieto (Eds), Springer, 2019.

\bibitem{Kin01}
A.H.E. Kinani, and M. Daoud, Generalized intelligent states for an arbitrary quantum system, \textit{J. Phys.A : Math. Gen} \textbf{43} (2001) 5373.

\bibitem{Dod02}
V.V. Dodonov, ``Nonclassical'' states in quantum optics: a ``squeezed'' review of the first 75 years, \textit{J. Opt. B Quantum Semiclass. Opt.} \textbf{4} (2002) R1.

\bibitem{Jef56}
H. Jeffreys and B. Swirles, \textit{Methods of Mathematical Physics} (3rd Edn.), Cambridge University Press, Cambridge, 1956.

\bibitem{Roy88}
H.L. Royden, \textit{Real Analysis} (3rd Edn.), Macmillan, New York, 1988.


\end{thebibliography}
\end{document}